\documentclass[prb,twocolumn,superscriptaddress,showpacs,amsmath,amssymb]{revtex4}
\usepackage{float}
\usepackage{amsmath}
\usepackage{bm}
\usepackage[usenames]{color}
\usepackage{verbatim}
\usepackage{graphicx}
\newcommand{\be}{\begin{equation}}
\newcommand{\ee}{\end{equation}}   
\newcommand{\bea}{\begin{eqnarray}}
\newcommand{\eea}{\end{eqnarray}}
\newcommand{\phrl}[1]{Phys.~Rev.~Lett. {\bf #1}}
\newcommand{\phrb}[1]{Phys.~Rev.~B {\bf #1}}
\newcommand{\phrx}[1]{Phys.~Rev.~X {\bf #1}}
\newcommand{\cmat}[1]{arXiv:{\bf #1}}

\newcommand{\jpcm}[1]{J.~Phys.:Condens.~Matter.{\bf #1}}
\newcommand{\bib}{\bibitem}
\newcommand{\lb}{\left[}
\newcommand{\rb}{\right]}
\newcommand{\lp}{\left(}
\newcommand{\rp}{\right)}
\newcommand{\lf}{\left\{}
\newcommand{\rf}{\right\}}

\renewcommand{\k}{\mathbf{k}}

\newcommand{\q}{\mathbf{q}}

\begin{document}
\title{Anomalous DC Hall response in noncentrosymmetric tilted Weyl semimetals}

\author{S. P. Mukherjee}
\affiliation{Department of Physics and Astronomy, McMaster University, Hamiltion, Ontario, Canada L8S 4M1}

\author{J. P. Carbotte}
\affiliation{Department of Physics and Astronomy, McMaster University, Hamiltion, Ontario, Canada L8S 4M1}
\affiliation{Canadian Institute for Advanced Research, Toronto, Ontario, Canada M5G 1Z8}

\begin{abstract}
Weyl nodes come in pairs of opposite chirality. For broken time reversal symmetry (TR) they are displaced in momentum space by $\bf{Q}$ and the anomalous DC Hall conductivity $\sigma_{xy}$ is proportional 
to $\bf{Q}$ at charge neutrality. For finite doping there are additive corrections to  $\sigma_{xy}$ which depend on the chemical potential as well as on the tilt ($C$) of the Dirac cones and on their 
relative orientation. If inversion symmetry (I) is also broken the Weyl nodes are shifted in energy by an amount $Q_{0}$. This introduces further changes in  $\sigma_{xy}$ and we provide simple analytic 
formulas for these modifications for both type I ($C<1$) and type II ($C>1$, overtilted) Weyl. For type I when the Weyl nodes have equal magnitude but oppositely directed tilts, the correction to 
$\sigma_{xy}$ is proportional to the chemical potential $\mu$ and completely independent of the energy shift $Q_{0}$. When instead the tilts are parallel, the correction is linear in $Q_{0}$ and $\mu$ 
drops out. For type II the corrections involve both $\mu$ and $Q_{0}$, are nonlinear and also involve a momentum cut off. We discuss the implied changes to the Nernst coefficient and to the thermal Hall 
effect of a finite $Q_{0}$.
\end{abstract}


\maketitle

\section{Introduction}
\label{sec:I}

Dirac semimetals are 3D analogues of 2D graphene first isolated in 2004.\cite{Novoselov} They harbor relativistic fermions with electronic dispersion curves which in the low energy sector are linear in 
all three momentum directions. The conduction and valence band meet at a single point and at charge neutrality their Fermi surface is point-like. Breaking either time-reversal (TR) or inversion symmetry 
can lift the two-fold degeneracy of a Dirac cone to produce a pair of Weyl nodes having opposite chirality. Broken TR-symmetry displaces the Dirac cones in momentum space by $\pm \bf{Q} $ while broken 
inversion symmetry displaces them in energy by $\pm Q_{0}$. Weyl nodes where first studied theoretically in pyrochlore iridates \cite{Savrasov,Ran} and other systems\cite{Balents,Rappe} and later 
suggested to exist in noncentrosymmetric transition-metal monophosphides.\cite{Dai} This theoretical prediction lead to the discovery of many such materials including TaAs,\cite{Ding,Xu,Lv,Yang} 
TaP \cite{Shi} and NbAs.\cite{Alidoust} An example of broken TR-symmetry is YbMnBi$_2$.\cite{Borisenko} Weyl semimetals (WSM) have many exotic properties.\cite{Hosur} Their surface states have open 
Fermi arcs which end at the projection on the surface of the bulk Weyl nodes of opposite chirality. In a magnetic field closed cyclotron orbits are possible which involve the arcs on opposite surfaces 
connected through the bulk Weyl nodes.\cite{Potter,Vishwanath,Analytis} Other anomalous properties include the chiral anomaly,\cite{Ninomiya,Adler,Neupane,Li,Ong,YBKim} a negative 
magnetoresistance,\cite{Huang,Spivak} Hall effect\cite{Burkov2,Burkov3} and other anomalous transport properties.\cite{Burkov,Burkov1,Hirschberger} The absorptive part of the AC longitudinal optical 
conductivity\cite{Chen,Sushkov,Neubauer,Xiao,Nicol}gives information on the bulk and reflects directly the relativistic, linear in momentum, dispersion curves of the Dirac fermions. In a 3D system this 
translates to a region of conductivity which is linear in photon energy.\cite{Nicol,Tabert} This is the analog of the constant interband background seen in graphene.\cite{Gusynin,Jiang} In more complicated 
cases \cite{Tabert} there can be more than one quasi-linear region.\cite{Sushkov} The chiral anomaly also manifests in optical absorption \cite{Ashby} which contains an image of the transfer of 
charge, in the presence of an external non orthogonal electric $\bf{E}$ and magnetic $\bf{B}$ field, from the node of one chirality to the other of the opposite chirality.

Another recent development is the theoretical realization that a new type of Weyl fermions can exist in condensed matter systems that is not part of high-energy physics and an example is WTe$_2$.\cite{Soluyanov} 
In this case the Dirac cones are tilted with respect to an axis in the Brillouin zone. If the tilt is small enough such that the electronic density of state at the node remains zero, the Weyl node is said 
to be type I but for large tilt, electron and hole pockets at zero energy can form as a result of a Lifshitz transition and this new phase is referred to as type II \cite{Soluyanov} with finite density of 
state at the Weyl point. Many properties of tilted Weyl cones have already been worked out, including its effect on the interband optical background,\cite{Carbotte, Chinotti} on the AC Hall conductivity
\cite{Steiner,Mukherjee} and on the absorption of circular polarized light.\cite{Mukherjee} Tilting leads to the squeezing of the Landau levels \cite{Tchoumakov,Yao} on application of an external magnetic 
field and even the collapse of the spectrum. New transitions appear in the optical spectrum beyond the dipolar ones and the surface Fermi arcs are modified.\cite{Goerbig} 

Central to the present work is the paper of Zyuzin and Tiwari [\onlinecite{Tiwari}] on the intrinsic anomalous Hall effect and very recent discussions of the Nernst effect \cite{Ferreiros,Saha} which 
apply to WSM when TR-symmetry has been broken. In this work we consider the DC Hall effect in type I and type II WSM when both TR-symmetry and inversion symmetry is broken.\cite{Zyuzin, Shan} This 
possibility has been discussed in R-Al-X family of compounds where R(rare earth), Al(aluminium) and X(Si, Ge).\cite{Singh} Our work also applies to broken inversion symmetry on its own and properly 
reduces to the results given in reference [\onlinecite{Tiwari}] when only TR-symmetry is broken. In section II we specify our continuum limit Hamiltion for a pair of Weyl nodes of opposite chirality 
displaced in momentum space by $\pm \bf{Q}$ and in energy by $\pm Q_{0}$. We work out a general formula for the DC limit of the Hall conductivity valid for a general tilt and chemical potential $\mu$. In 
section III we consider the limit of zero chemical potential ($\mu=0$) for both the case when the Dirac cones are oppositely tilted (inversion symmetric tilt) and when they are tilted in the same direction 
(tilt violates inversion symmetry). Our final analytic expressions properly reduce to those of reference [\onlinecite{Tiwari}] when $Q_{0}$ is taken to be zero. In section IV we consider the finite $\mu$ 
case. Our result reproduce those found in section III when we take $\mu=0$. This requires we employ the expressions obtained under the assumption that $\mu<Q_{0}$. Different expressions apply when $\mu>Q_{0}$. 
This second set of results are used to show that when $Q_{0}$ is set to zero we reproduce the finite $\mu$ case with only broken TR-symmetry. We also discuss the case of broken inversion symmetry retaining TR. 
Section V deals with the Nernst effect. Again the generalized expression we obtain reduce to those described in recent literature \cite{Ferreiros,Saha} for the case of only broken TR-symmetry. A discussion 
and conclusions are given in section VI.

\section{Formalism}
\label{sec:II}

\begin{figure}
\centering
\includegraphics[width=1.5in,height=1.8in, angle=0]{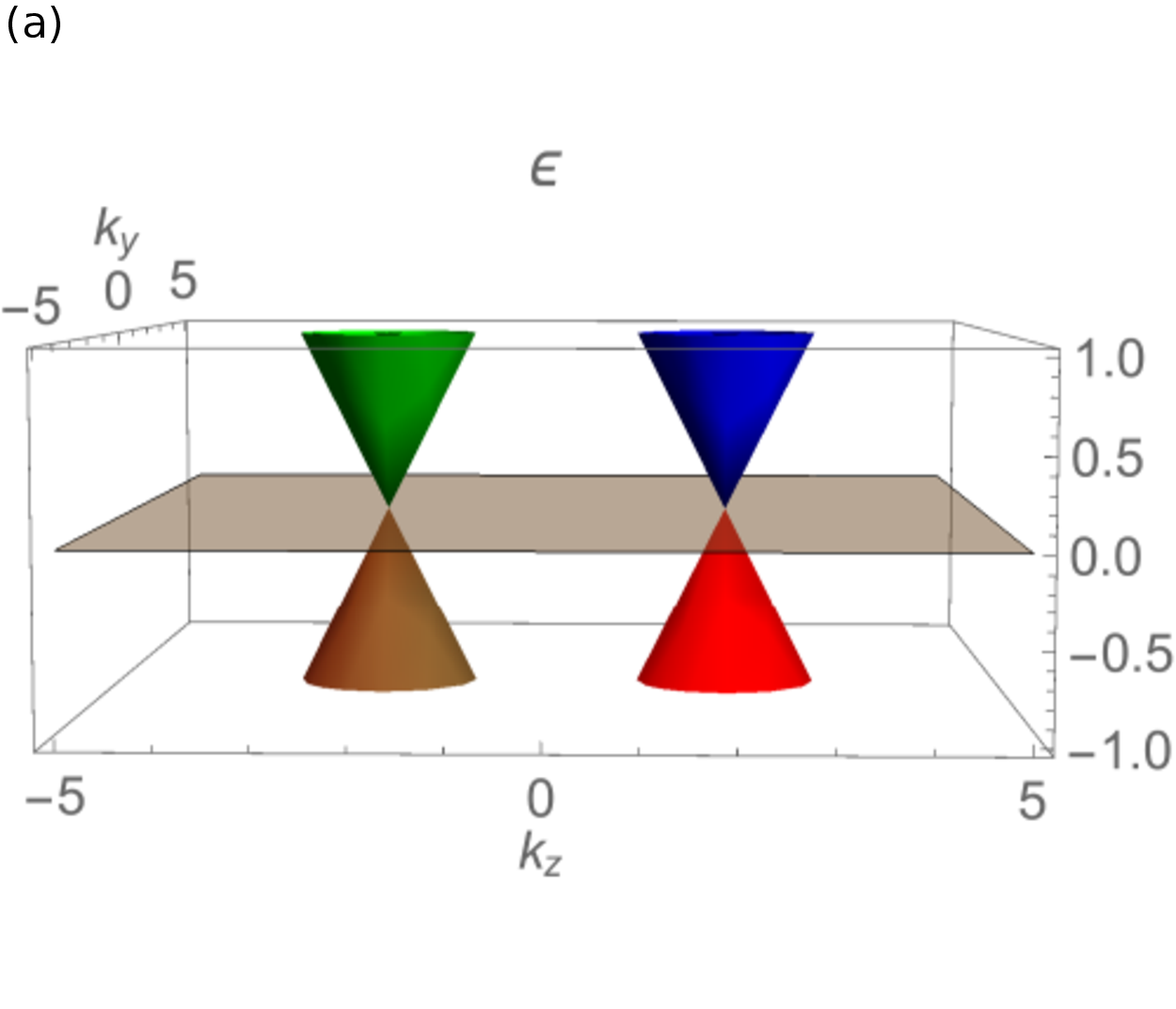}
\includegraphics[width=1.5in,height=1.8in, angle=0]{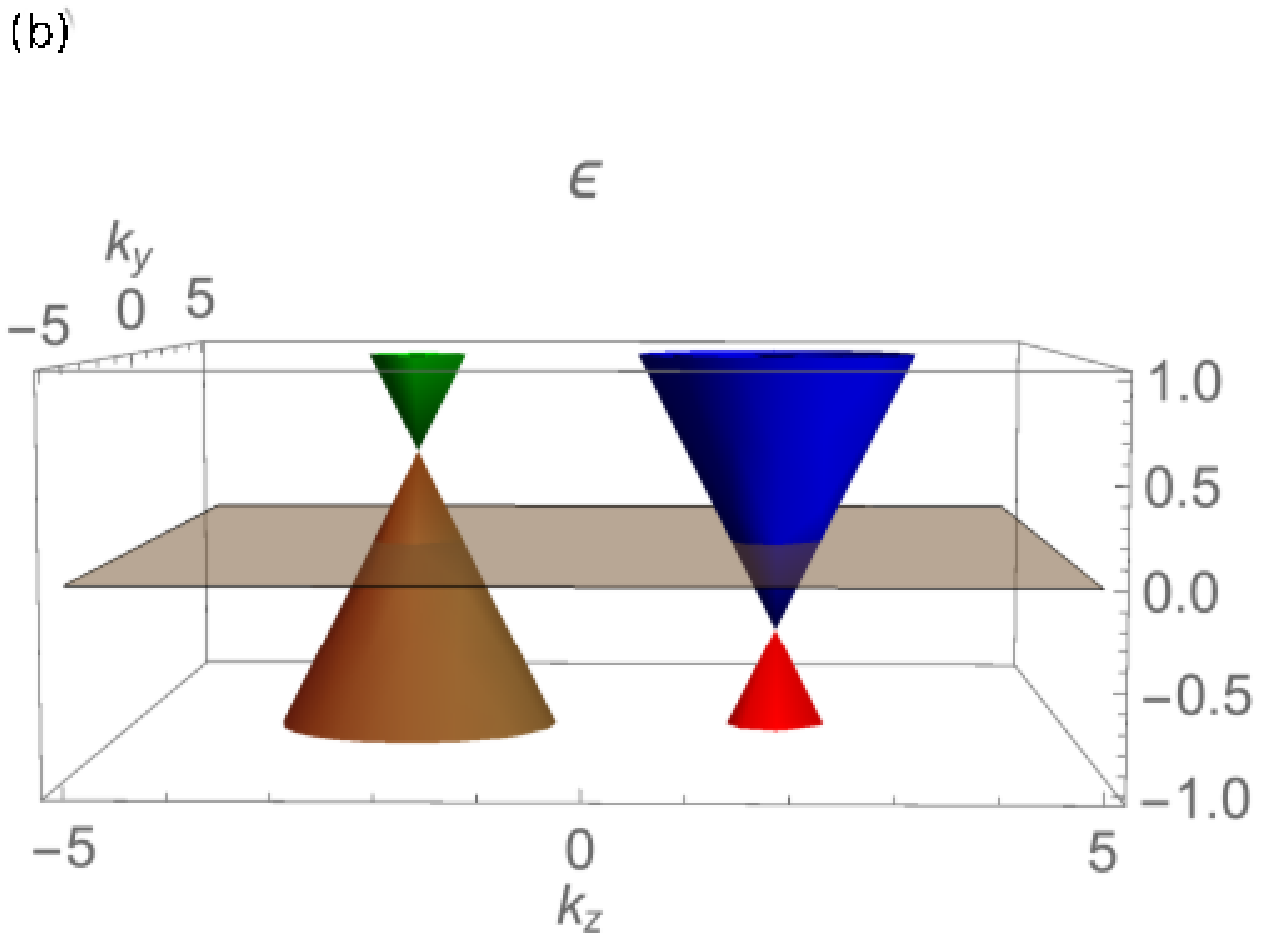}\\
\vspace{-0.9cm}
\includegraphics[width=1.5in,height=1.8in, angle=0]{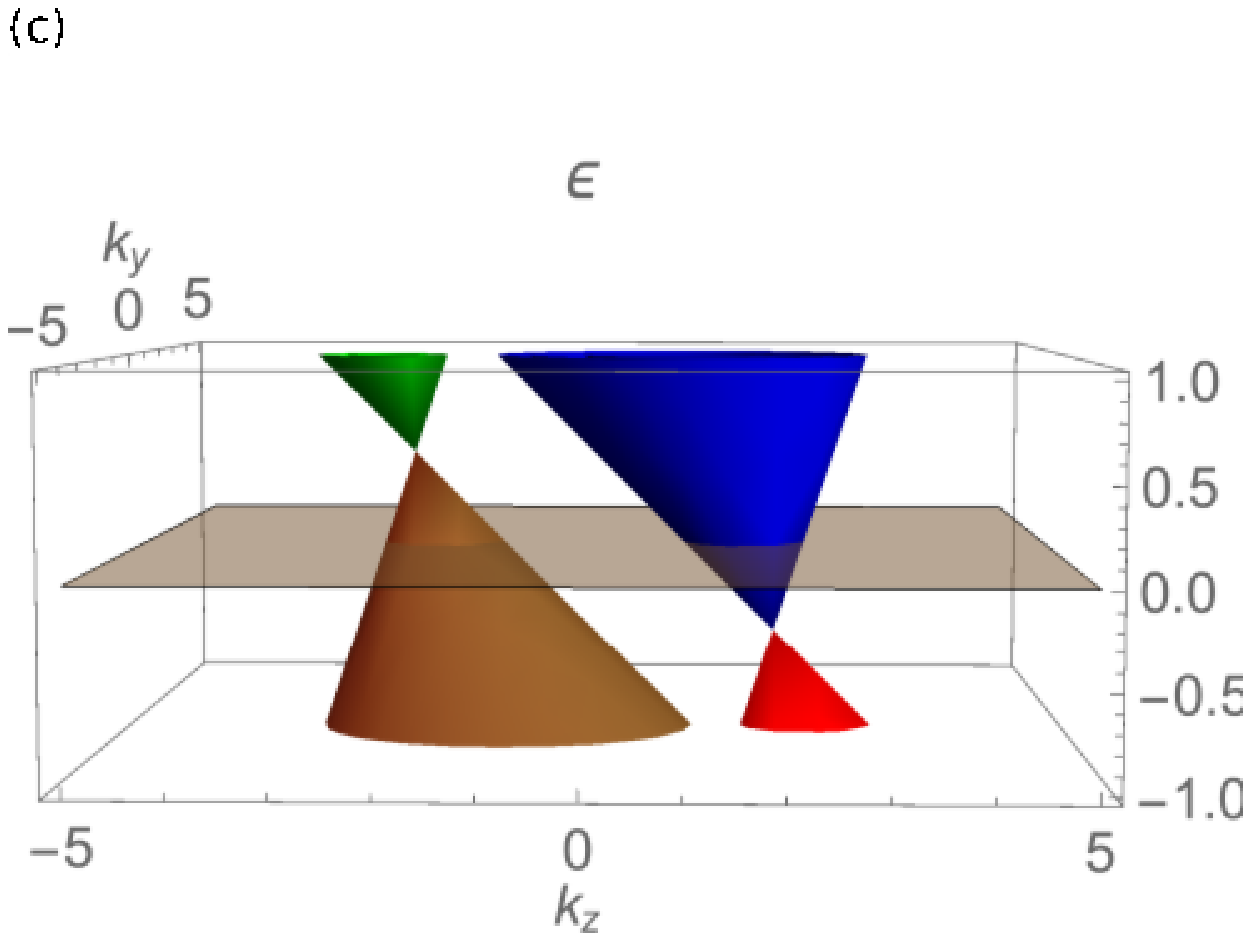} 
\includegraphics[width=1.5in,height=1.8in, angle=0]{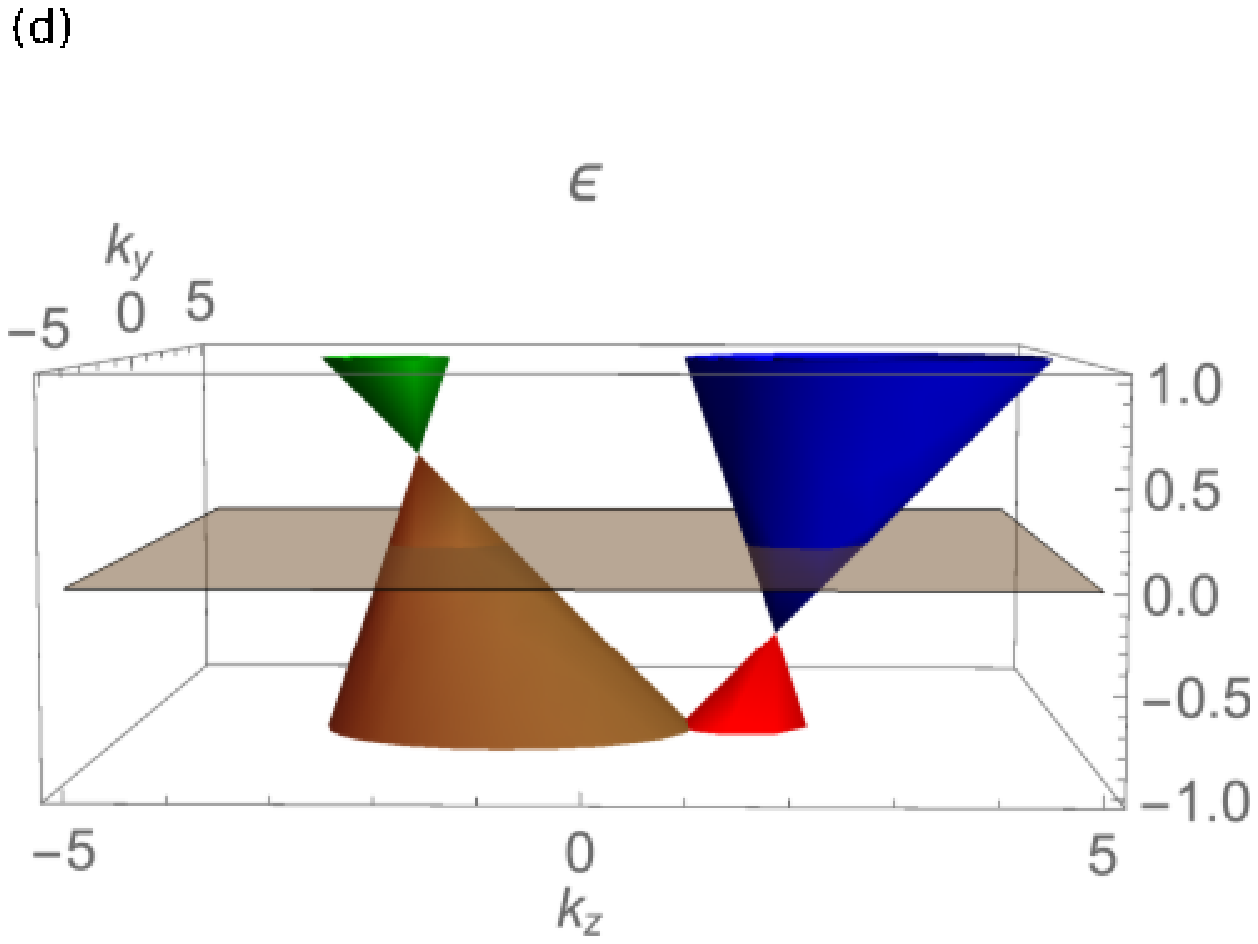} 
\caption{(Color online)The actual 3D plot of the energy dispersion showing the orientations of the Weyl cones close to the Fermi energy. Here we plot the energy dispersion as represented in 
Eq.(\ref{Dispersion}).The blue and red cones are respectively the conduction band ($s=+$) and valence band ($s=-$) of the Weyl node with positive chirality ($s'=1$). Similarly green and brown cones 
are respectively the conduction band ($s=+$) and valence band ($s=-$) of the Weyl node with negative chirality ($s'=-1$).The two chiral Weyl nodes are separated by a distance $\pm Q$. The gray shaded 
plane corresponding to the $\mu=0$ plane. We set $v=1$. $Q$ is taken twice the Fermi velocity and $Q_{0}$ half of it. (a) Two untilted Weyl nodes separated by $Q$ in the inversion symmetry preserving 
case. (b) Two untilted Weyl nodes after we break the inversion symmetry. (c) Two tilted Weyl nodes with $C_{1}=C_{2}=0.5$ and $Q_{0}=0.5$. (d) Same two Weyl cones as in in (c) but oppositely tilted 
($C_{1}= - C_{2} = -0.5 $).}
\label{Fig1}
\end{figure}

We begin with the minimal continuum Hamiltion for a pair of Weyl node of opposite chirality with both TR and inversion symmetry broken. The first displaces the Dirac cone in momentum space by an amount 
$\pm \bf{Q}$ while the second shifts their energy by $\pm Q_{0}$. The Hamiltion is given by the following equation,\cite{Nicol,Zyuzin,Shan}
\be
\hat{H}_{s'}(\k)=C_{1,2}(k_{z}-s' Q)+s' v \bm{\sigma}.(\k-s'Q\bm{e}_z)-s'Q_{0}
\label{Hamiltonian}
\ee
where $s'=1$ for Weyl point indexed by 1 and  $s'=-1$ for Weyl point indexed by 2. $C_{1,2}$ describe the amount of tilting of the particular chiral node, $v$ the Fermi velocity and $\bm{e}_{i}$ the 
unit vector along the axis $x_{i}$ where $i=x, y, z$. The Pauli matrices are defined as usually by,
\be
\sigma_{x}=\lp\begin{array}{cc}0 & 1\\ 1 & 0 \end{array} \rp, \sigma_{y}=\lp\begin{array}{cc}0 & -\imath\\ \imath & 0 \end{array} \rp, \sigma_{z}=\lp\begin{array}{cc}1 & 0\\ 0 & -1 \end{array} \rp.
\ee
The broken inversion symmetry is introduced through the third term in the Hamiltonian. The energy dispersion corresponding to the above Hamiltonian is described as,
\bea
&& \hspace{-0.2cm}\epsilon_{s,s'}\hspace{-0.1cm}=\hspace{-0.1cm}C_{1,2}k_{z}\hspace{-0.1cm}-\hspace{-0.1cm}s'C_{1,2}Q\hspace{-0.1cm}-\hspace{-0.1cm}s'Q_{0}\hspace{-0.1cm}+\hspace{-0.1cm}sv\sqrt{k^2-\hspace{-0.1cm}2s'k_{z}Q\hspace{-0.1cm}+\hspace{-0.1cm}Q^{2}}
\label{Dispersion}
\eea
where $s=\pm$ stands for conduction($+$) and valence($-$) bands and $k^2=k^2_{x}+k^2_{y}+k^2_{z}$. For a set of values of the parameters ($v=1, Q=2, Q_{0}=0.5$) we plot in Fig.(\ref{Fig1}) the energy 
dispersion for the different cases of tilting. When we take the inversion symmetry breaking into account the negative chiral Weyl node gets shifted upward by an amount $Q_{0}=0.5$ while the opposite 
happens for the other chiral node. 

\begin{figure}
\includegraphics[width=1.5in,height=1.5in, angle=0]{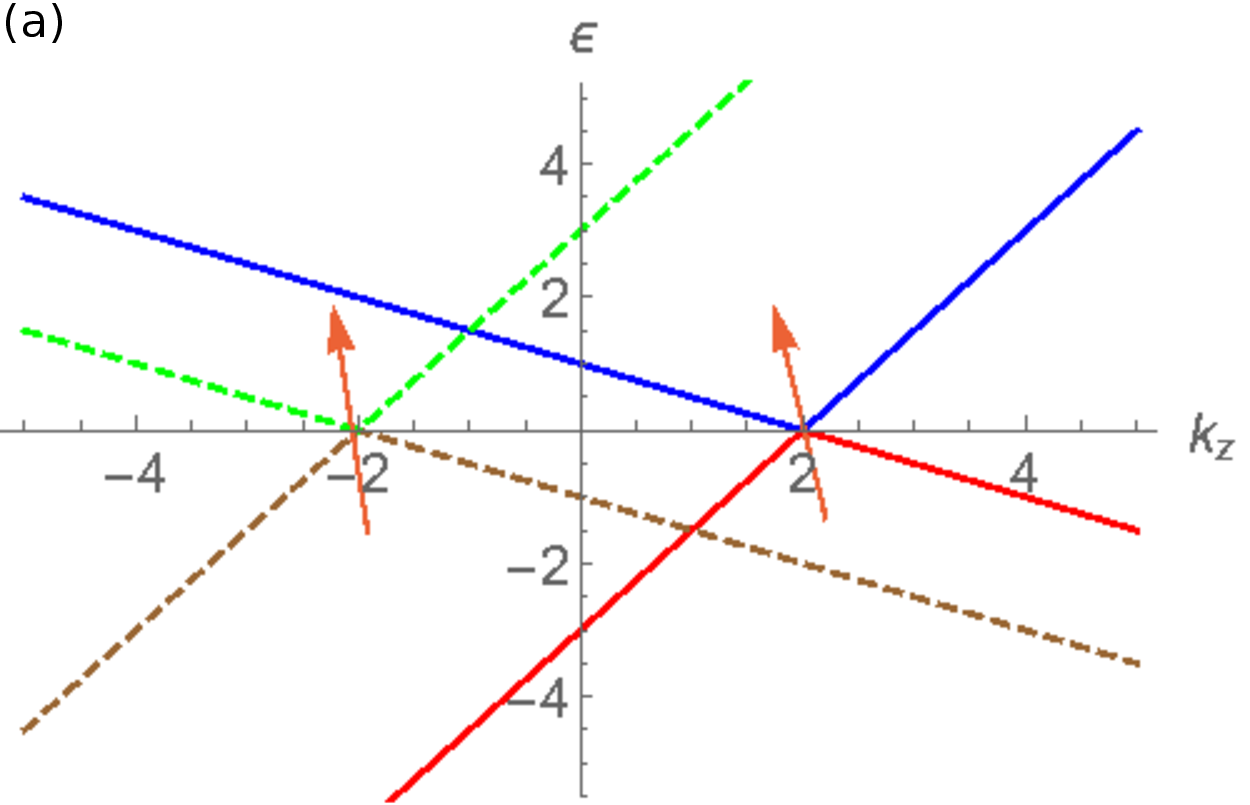}
\includegraphics[width=1.5in,height=1.5in, angle=0]{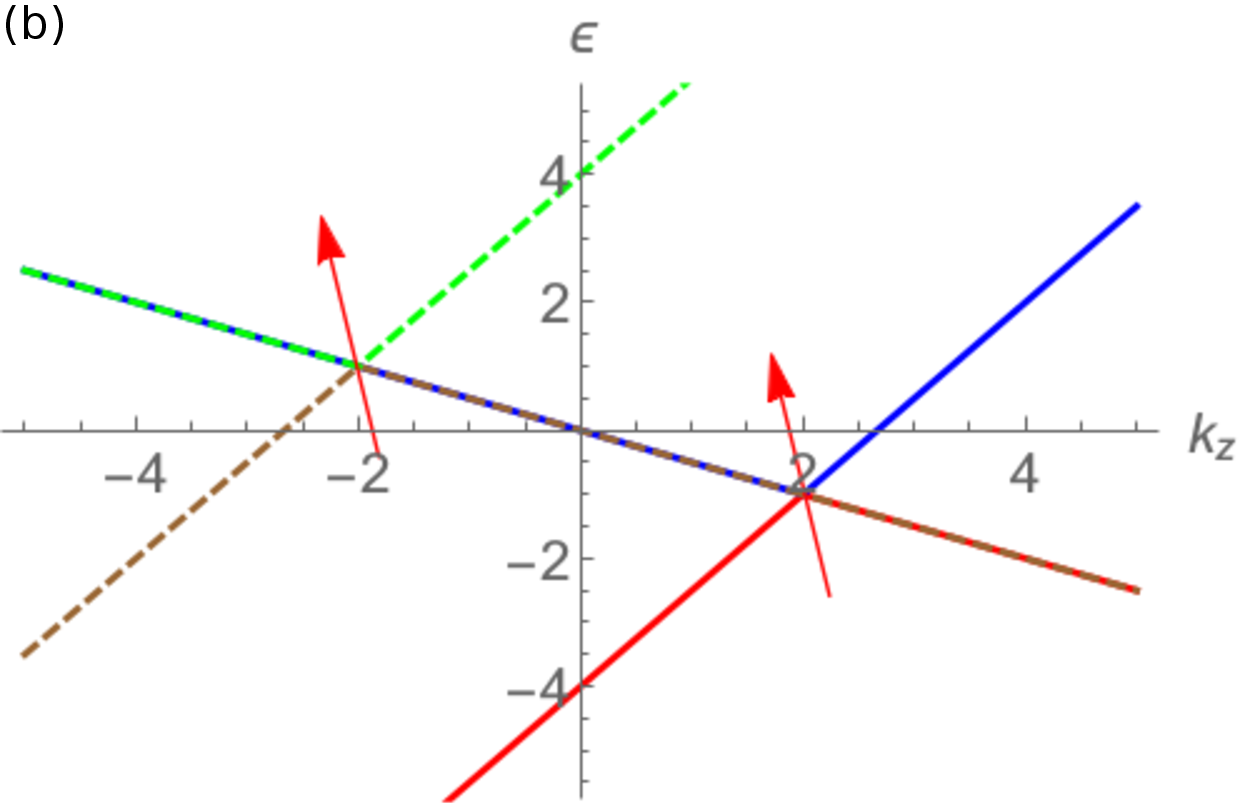}\\
\includegraphics[width=1.5in,height=1.5in, angle=0]{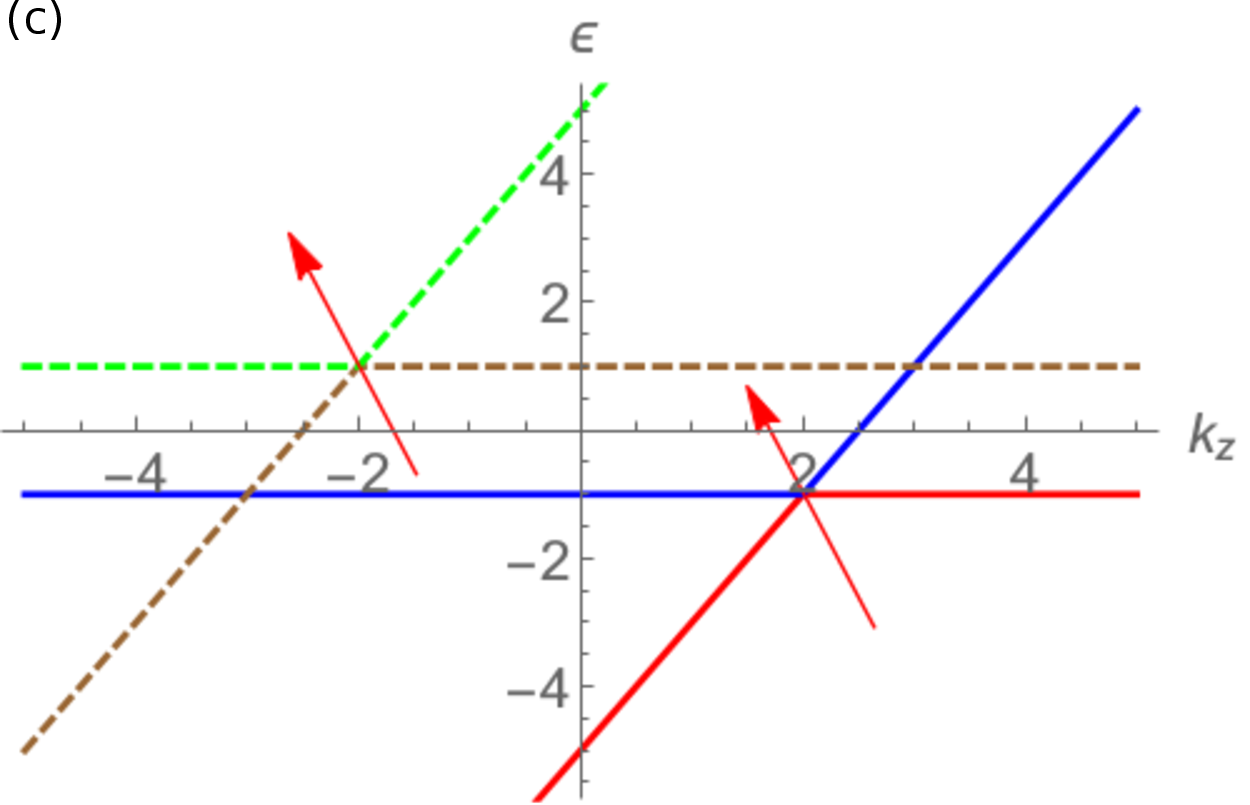} 
\includegraphics[width=1.5in,height=1.5in, angle=0]{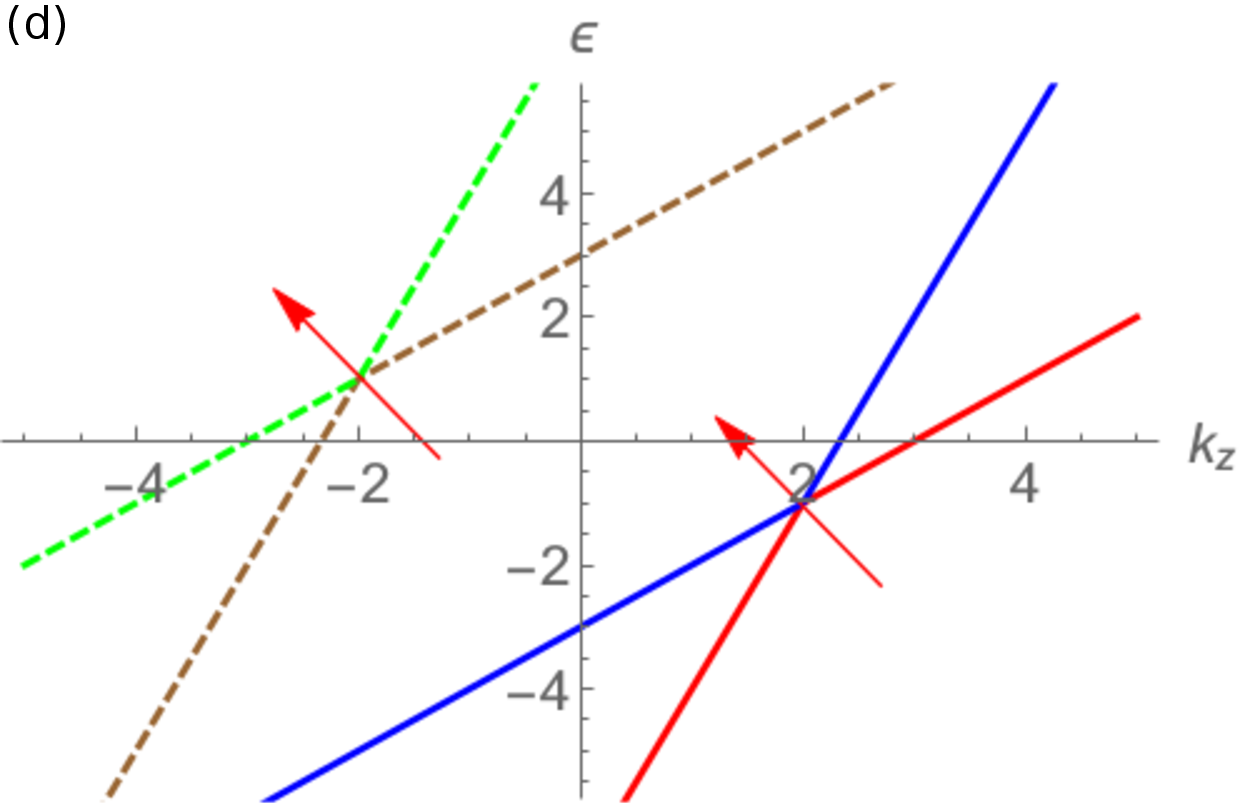} 
\caption{(Color online) Evolution of parallel tilted Weyl cones as we change the amount of the tilt. Here we follow the same color code used in Fig.\ref{Fig1}. We also keep the value of the parameters 
$v,Q,Q_{0}$ same. Additionally we set $k_{x}=k_{y}=0$ and draw the outlines of the positive chiral Weyl nodes by solid lines and that for the negative one with broken lines. Fig. (a) shows the two Weyl 
nodes with the tilt $C_{1}=C_{2}=0.5$ for inversion symmetric system. Fig.(b) shows the same tilted Weyl nodes with broken inversion symmetry. In Fig.(c) we show the case for the tilting $C_{1}=C_{2}=1$ 
and in (d) for $C_{1}=C_{2}=2$. With increasing tilt the Weyl cones also gradually become wider.} 
\label{Fig2}
\end{figure}

\begin{figure}
\includegraphics[width=1.5in,height=1.5in, angle=0]{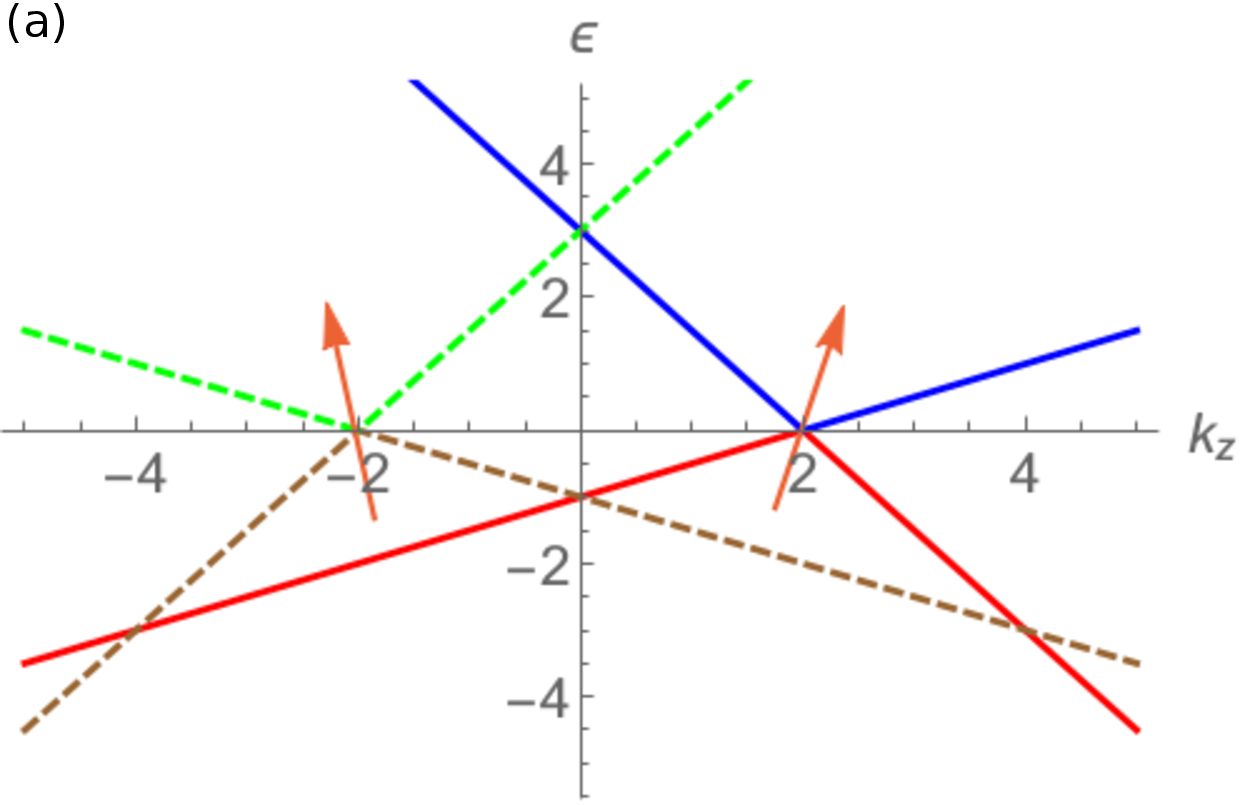}
\includegraphics[width=1.5in,height=1.5in, angle=0]{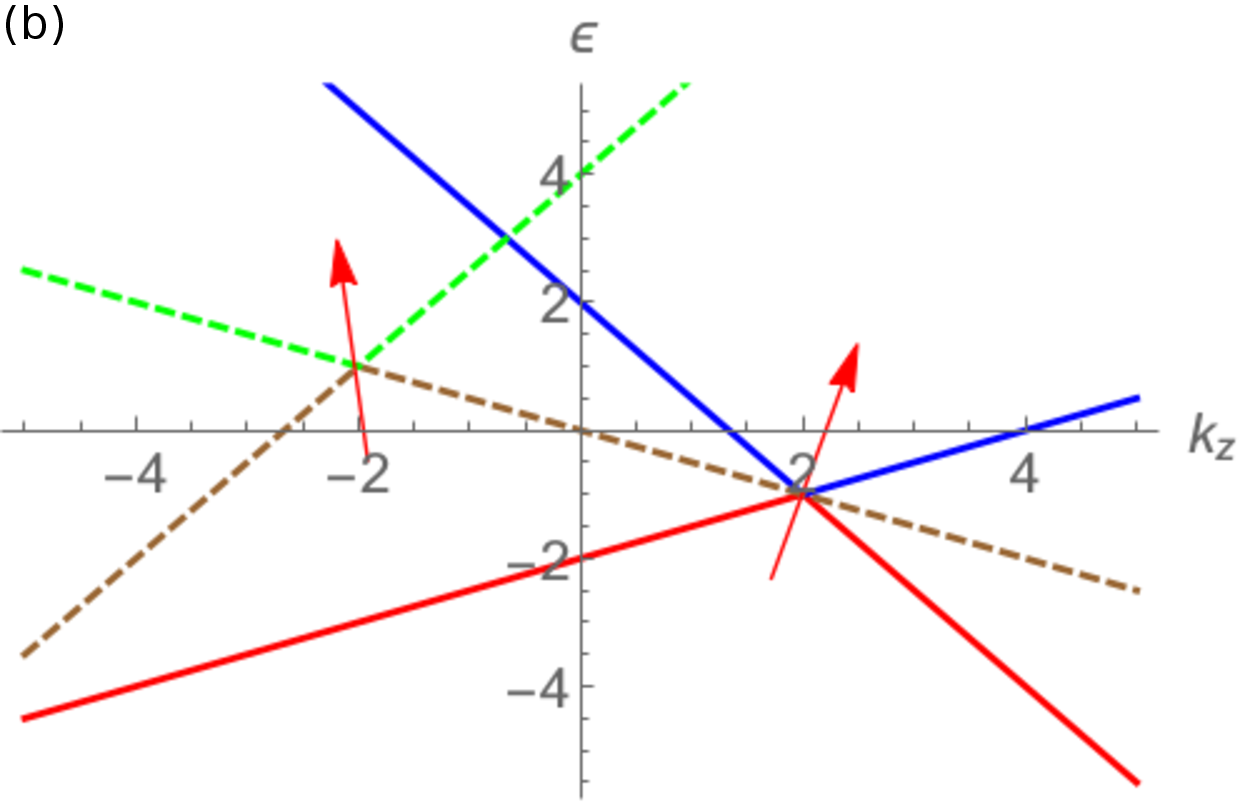} \\
\includegraphics[width=1.5in,height=1.5in, angle=0]{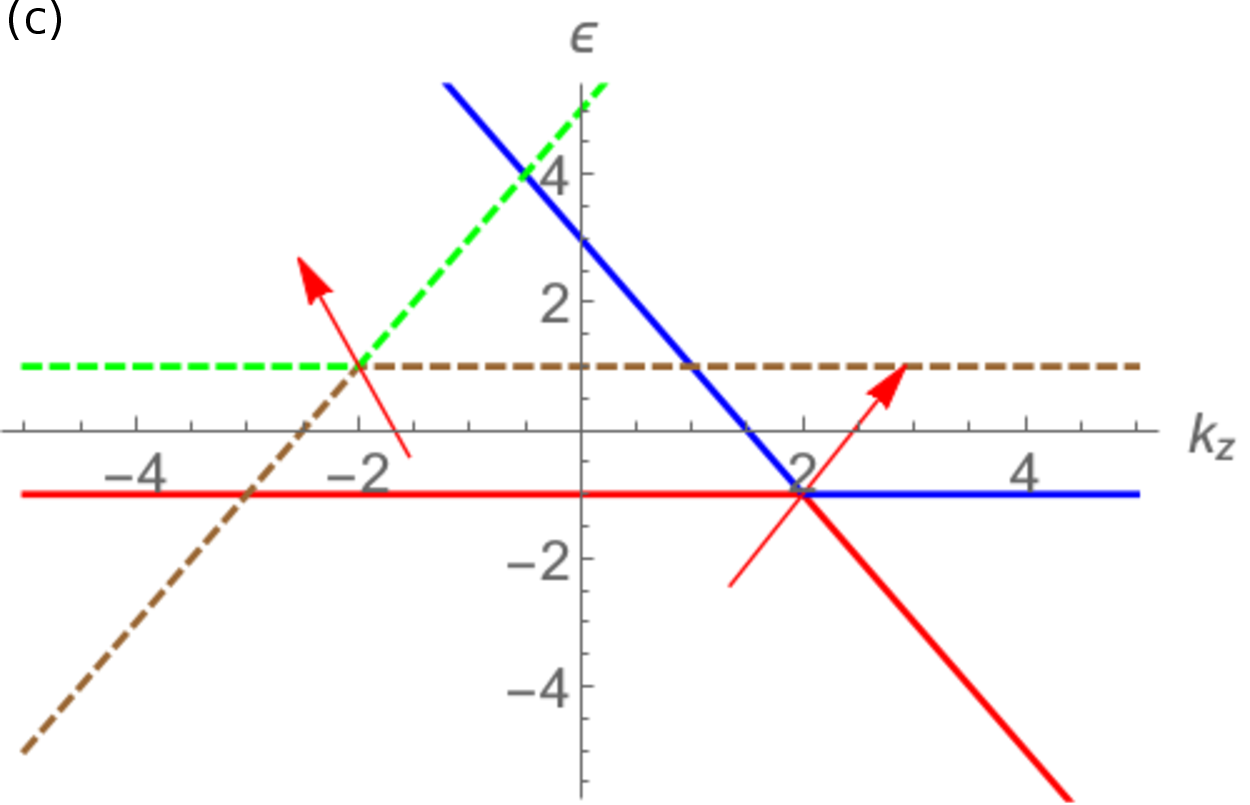} 
\includegraphics[width=1.5in,height=1.5in, angle=0]{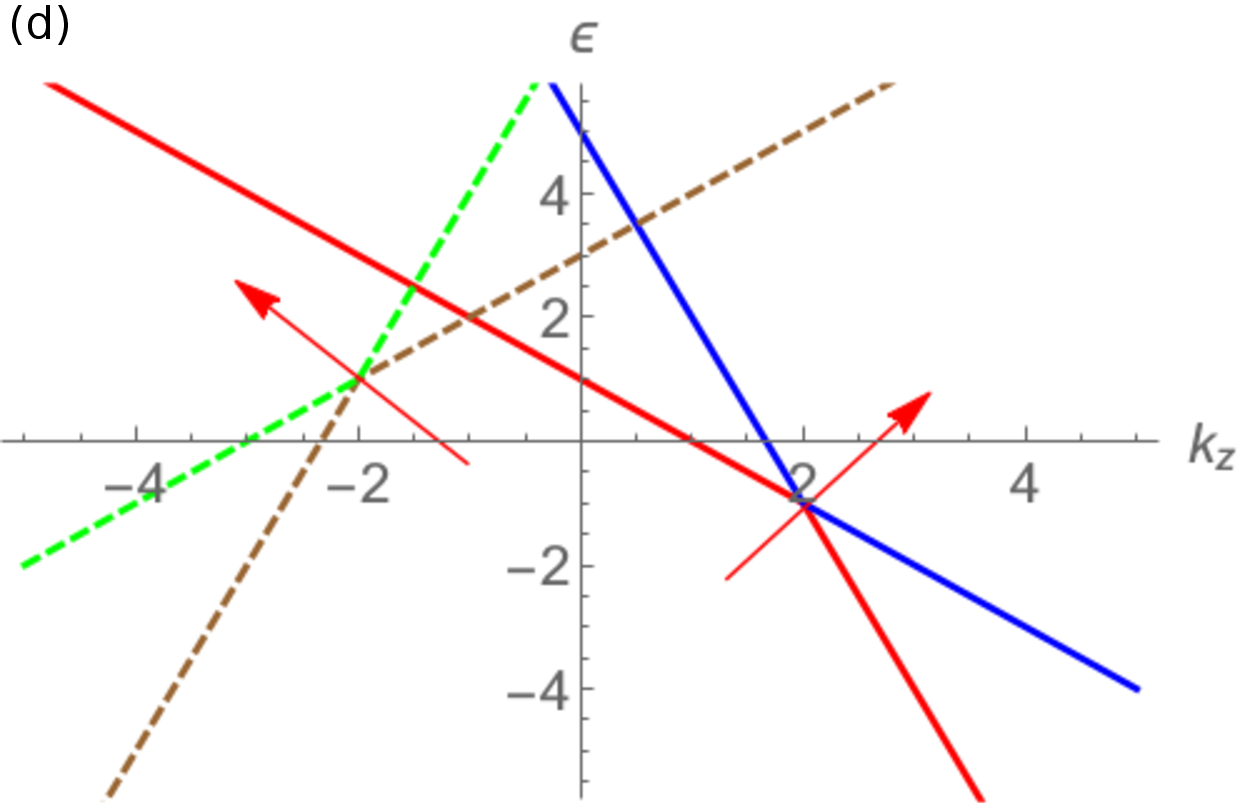} 
\caption{(Color online) Evolution of two oppositely tilted ($C_{1}=-C_{2}$) Weyl nodes as we change the amount of the tilt. Here we follow the same color code and the parameter set as in Fig.(\ref{Fig2}). 
Fig.(a) shows the two Weyl nodes with the tilt $C_{1}=-C_{2}=-0.5$ for inversion symmetric system. Fig.(b) shows the same tilted Weyl cones with broken inversion symmetry. In Fig.(c) we show the case for 
the tilting when $C_{1}=-C_{2}=-1$ and in (d) for $C_{1}=-C_{2}=-2$.}
\label{Fig3}
\end{figure}

\begin{figure}
\includegraphics[width=1.6in,height=2.0in, angle=0]{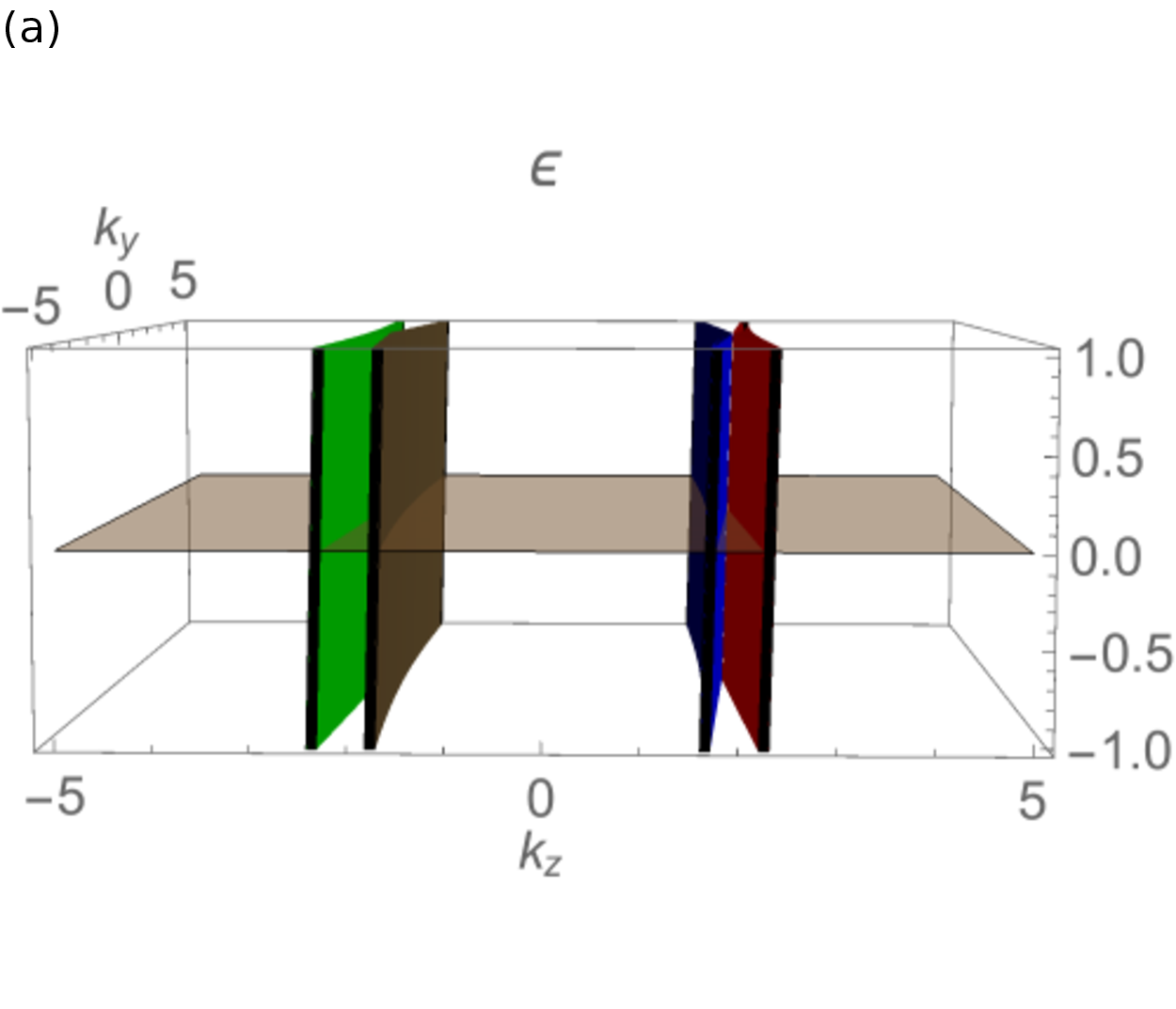} 
\includegraphics[width=1.6in,height=2.0in, angle=0]{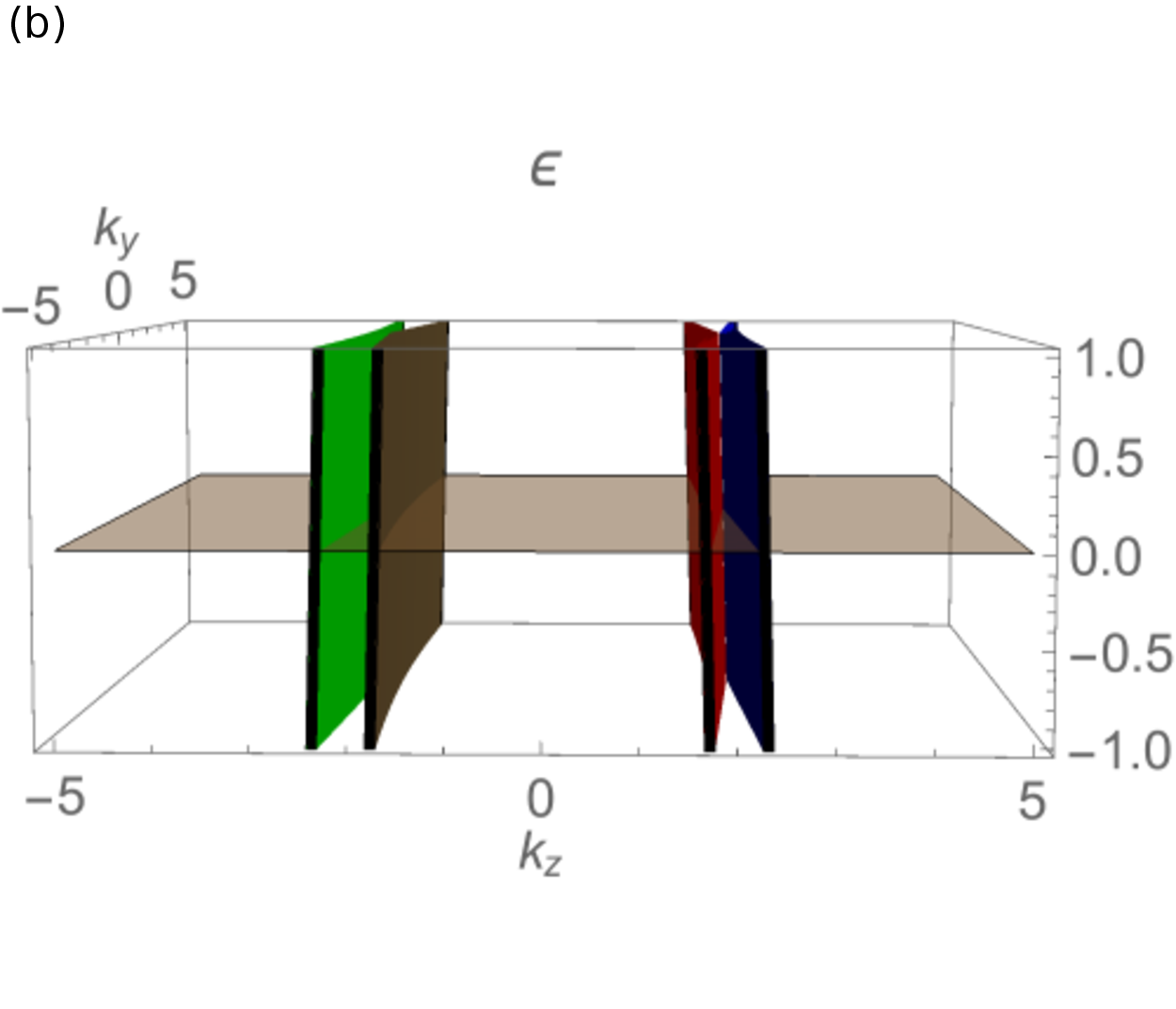} 
\caption{(Color online) 3D diagram showing final stage of evolution of the Weyl nodes. As we increase the tilt the Weyl cones open up more and more and ultimately they merge with two sets of planes 
displaced by $Q$. In Fig.(a) we show the case of large tilting for the parallel tilted ($C_{1}=C_{2}=20$) Weyl cones and in Fig.(b) we show the same for two oppositely tilted ($C_{1}=-C_{2}=-20$) Weyl 
cones. Note that the order of the blue and red planes has switched in frame (b) as compared with (a).}
\label{Fig4}
\end{figure}

In Fig.(\ref{Fig2}) we study the evolution of the Weyl nodes when they are tilted parallel to each other for the case $C_{1}=C_{2}$ for different amount of tilting. We see that as we increase tilting 
individual Weyl nodes not only bend but also become progressively wider and ultimately the conduction and valence bands merge with the planes $\epsilon_{s,s'}=C_{1,2}(k_{z}-s'Q)$ \cite{Tiwari} for 
large tilting (see Fig.(\ref{Fig4}a)). Similarly, we consider the cases of opposite tilting ($C_{1}=-C_{2}$) in Fig.(\ref{Fig3}) and we can also draw the same inference as shown in Fig.(\ref{Fig4}b).

The Green's function corresponding to the above Hamiltonian is given by,
\be
G_{s'}(k,z)= \lb I_{2}z-\hat{H}_{s'}(\k)\rb^{-1},
\label{GF-definition}
\ee
where $I_{2}$ is a $2\times 2$ unit matrix. It is straight forward to show that the Green's function can be written in the following form,
\be
G_{1,2}(k,\imath \omega_{n})\hspace{-0.1 cm}=\hspace{-0.2 cm}\sum_{s=\pm}\hspace{-0.15cm} \frac{1-ss'\bm{\sigma}.\bm{N}_{\k -s'Q \bm{e}_{z}}}{\imath \omega_{n}\hspace{-0.1 cm}-\hspace{-0.1 cm}C_{1,2}(k_{z}-s' Q)\hspace{-0.1 cm}+\hspace{-0.1 cm}sv|\k\hspace{-0.1 cm}-\hspace{-0.1 cm}s' Q\bm{e}_z|\hspace{-0.1 cm}+\hspace{-0.1 cm}s'Q_{0}},
\label{GF-full}
\ee
where $\bm{N}_{\k-s'Q \bm{e}_{z}}=\frac{k_{x}\bm{e}_{x}+k_{y}\bm{e}_{y}+(k_{z} -s'Q)\bm{e}_{z}}{\sqrt{k^2_{x}+k^2_{y}+(k_{z}-s' Q)^2}}$.

Since in the subsequent sections we will discuss the behavior of anomalous Hall conductivity $\sigma_{xy}$, we need the corresponding current-current correlation function within the realm of the Kubo 
formalism which is defined as,
\bea
&& \hspace{-0.5 cm}\Pi_{xy}(\Omega_{m},\q)=\hspace{-0.1 cm}T\sum_{\omega_{n}}\sum_{s'=\pm}\hspace{-0.1 cm} \int \frac{d^3k}{(2\pi)^3} J_{x,s'}G_{1,2}(\k+\q,\omega_{n}+\Omega_{m}) \nonumber\\
&& \times J_{y,s'}G_{1,2}(\k,\omega_{n})\nonumber\\
&& =Te^2v^2\sum_{\omega_{n}}\sum_{s'=\pm} \int \frac{d^3k}{(2\pi)^3} \sigma_{x}G_{1,2}(\k+\q,\omega_{n}+\Omega_{m})\times \nonumber\\
&& \sigma_{y}G_{1,2}(\k,\omega_{n}),
\eea
where the sum $\omega_{n}$ is over the Fermionic Matsubara frequencies and $\Omega_{m}$ is an external Bosonic Matsubara frequency. We have used the definition of the current operators,
\be
J_{\{x,y\},s'}=s'ev\sigma_{\{x,y\}}.
\ee
With these definitions we calculate the expression for the correlation function after setting $\q$ to zero as,
\bea
&& \Pi_{xy}(\Omega_{m},0)=e^2 \sum_{s'=\pm}s'\int^{\Lambda-s'Q}_{-\Lambda-s'Q} \frac{dk_{z}}{2\pi}\int^{\infty}_{0} \frac{k_{\perp}dk_{\perp}}{2\pi}\times \nonumber \\ 
&& \{f(C_{s'}k_{z}+vk-s'Q_{0})-f(C_{s'}k_{z}-vk-s'Q_{0})\} \times \nonumber \\  
&& \frac{k_{z}}{k}\lb \frac{2v^2\Omega_{m}}{\Omega^2_{m}+4v^2k^2}\rb.
\label{Correlation-func}
\eea
Here $\Omega_{m}$ is the Matsubara frequency, $\Lambda$ the cutoff, $k_{\perp}$ is the momentum perpendicular to $k_{z}$ and $f(E)=(e^{(E-\mu)/T}+1)^{-1}$ is the Fermi function at finite temperature 
$T$ with $\mu$ the chemical potential. Replacing $\imath\Omega_{m}$ with $\Omega+\imath \delta$, the DC conductivity is,
\bea
&& \sigma_{xy}=-\lim_{\Omega \to 0} \frac{\Pi_{xy}(\Omega,0)}{\imath\Omega}=\frac{e^2 v^2}{2\pi^2} \sum_{s'=\pm}s'\int^{\Lambda-s'Q}_{-\Lambda-s'Q} k_{z} dk_{z} \nonumber\\ 
&& \int^{\infty}_{0} \frac{k_{\perp}dk_{\perp}}{k} \biggl\{f(C_{s'}k_{z}+vk-s'Q_{0})-\nonumber\\ 
&& f(C_{s'}k_{z}-vk-s'Q_{0})\biggr\} \lb\frac{1}{4v^2k^2}+\imath\pi \delta(4v^2k^2)\rb.
\eea

We write the real part of the Hall conductivity as,
\bea
&& \hspace{-0.2cm}\Re {\sigma_{xy}}=\frac{e^2}{8\pi^2} \sum_{s'=\pm}s'\int^{\Lambda-s'Q}_{-\Lambda-s'Q} k_{z} dk_{z} \int^{\infty}_{|k_{z}|} \frac{dk}{k^2} \times \nonumber\\
&& \hspace{-0.2cm}\{f(C_{s'}k_{z}+vk-s'Q_{0})-f(C_{s'}k_{z}-vk-s'Q_{0})\} ,
\eea
where we have replaced $k_{\perp}$ with $k$ following the relation $k=\sqrt{k^2_{\perp}+k^{2}_{z}}$. Now we let the temperature $T$ go to zero and replace the Fermi functions appropriately with the 
Heaviside step functions $\Theta$ as shown below,
\bea
&& \Re {\sigma_{xy}}=-\frac{e^2}{8\pi^2} \sum_{s'=\pm}s'\int^{\Lambda-s'Q}_{-\Lambda-s'Q} dk_{z} \nonumber \\ 
&& \biggl [\text{sgn}(k_{z})\Theta(v^2k^2_{z}-(C_{s'}k_{z}-\mu_{s'})^2)+\frac{vk_{z}}{|C_{s'}k_{z}-\mu_{s'}|} \nonumber\\
&& \lp1-\Theta(v^2 k^2_{z}-(C_{s'}k_{z}-\mu_{s'})^2)\rp\biggr],
\label{AHC}
\eea
where $\mu_{s'}=\mu+s'Q_{0}$ is the chemical potential for the Weyl node with chirality $s'$ and 'sgn' is the sign function. From the above equation we see that the sole effect of the inversion symmetry 
breaking is coming through the Fermi function where the chemical potential $\mu$ is getting replaced by $\mu_{s'}$. We will use this equation for the anomalous Hall conductivity in the subsequent sections
of this article. 

\section{Anomalous Hall conductivity at charge neutrality}
\label{sec:III}

In this section we discuss the behavior of anomalous Hall conductivity, as described in the last section at charge neutrality. For this we set $\mu=0$ in Eq.(\ref{AHC}) and write it as,

\bea
&& \Re {\sigma_{xy}}=-\frac{e^2}{8\pi^2} \sum_{s'=\pm}s'\int^{\Lambda-s'Q}_{-\Lambda-s'Q} dk_{z} \nonumber \\
&& \biggl[ \text{sgn}(k_{z})\Theta(v^2k^2_{z}-(C_{s'}k_{z}-s'Q_{0})^2)+\frac{vk_{z}}{|C_{s'}k_{z}-s'Q_{0}|}  \nonumber \\ 
&& \times \lp1-\Theta(v^2 k^2_{z}-(C_{s'}k_{z}-s'Q_{0})^2)\rp\biggr]. 
\eea
We can divide the above integral into positive and negative range of $k_{z}$ and make the variable transformation $k_{z}\to -k_{z}$ in the negative part. Further we see that under the interchange of the 
chirality ($s' \to -s'$) the total contribution of the two integrals stays the same. This allows us to drop the $s'$ index except in the node specific tilt term $C'_{s'}$. 
\bea
&& \hspace{-0.4cm}\Re {\sigma_{xy}}=-\frac{e^2}{8\pi^2} \sum_{s'=\pm}\biggl[\int^{\Lambda-Q}_{0} dk_{z} \biggl\{\Theta(k^2_{z}-(C'_{s'}k_{z}-Q'_{0})^2)\nonumber\\
&& +\frac{k_{z}}{|C'_{s'}k_{z}-Q'_{0}|} \lp1-\Theta(k^2_{z}-(C'_{s'}k_{z}-Q'_{0})^2)\rp\biggr\}-\nonumber\\
&&\int^{\Lambda+Q}_{0} dk_{z} \biggl\{\Theta(k^2_{z}-(C'_{s'}k_{z}+Q'_{0})^2)+\frac{k_{z}}{|C'_{s'}k_{z}+Q'_{0}|} \nonumber \\ 
&& \times \lp1-\Theta(k^2_{z}-(C'_{s'}k_{z}+Q'_{0})^2)\rp\biggr\}\biggr].
\label{AHC1}
\eea
Here $C'_{s'}$ denotes the anti-clockwise tilt for the Weyl node with chirality $s'$, normalized by the Fermi velocity $v$ and is a dimensionless parameter in our derivation. Similarly $Q'_{0}=Q_{0}/v$ 
and having the dimension of momentum.

We can differentiate between the two cases depending on whether $C'_{s'}$ is less than (WSM type I) or greater than one (WSM type II). First we see what happens when $C'_{s'}<1$. From the above Eq.(\ref{AHC1}) 
we see that for $C'_{s'}<1$ the argument of the $\Theta$ function is positive only when $\Lambda-Q>k_{z}>\frac{Q'_{0}}{1+C'_{s'}}$ (in the first integral) and $\Lambda+Q>k_{z}>k_{z}>\frac{Q'_{0}}{1-C'_{s'}}$ 
(in the second integral). In the range $k_{z}<\frac{Q'_{0}}{1+C'_{s'}}$, the quantity $|C'_{s'}k_{z}-Q'_{0}|$ is $Q'_{0}-C'_{s'}k_{z}$. Separating the contributions from two different Weyl nodes to the 
conductivity $\Re{\sigma_{xy}}$, we write the individual contribution from $s'$ node as $\Re {\sigma^{I}_{xy,s'}}$ which is given by,
\bea
&& \Re {\sigma^{I}_{xy,s'}}=\frac{e^2}{8\pi^2} \biggl[ \int^{\frac{Q'_{0}}{1-C'_{s'}}}_{0} \frac{k_{z}dk_{z}}{C'_{s'}k_{z}+Q'_{0}}+ \int^{\Lambda+Q}_{\frac{Q'_{0}}{1-C'_{s'}}} dk_{z}  \nonumber\\
&& -\int^{\frac{Q'_{0}}{1+C'_{s'}}}_{0} \frac{k_{z}dk_{z}}{Q'_{0}-C'_{s'}k_{z}}- \int^{\Lambda-Q}_{\frac{Q'_{0}}{1+C'_{s'}}} dk_{z} \biggr] .
\eea
After evaluating the integrals we get, 
\bea
&& \hspace{-0.6cm}\frac{\Re {\sigma^{I}_{xy,s'}}}{e^2 Q/2\pi^2}=\frac{1}{2}+\frac{1}{2}\biggl\{\frac{1}{C'_{s'}}+\frac{1}{2C'^{2}_{s'}}\ln\lp \frac{1-C'_{s'}}{1+C'_{s'}}\rp\biggr\}\frac{Q'_{0}}{Q}.
\label{HallI-anti-clock}
\eea 
Next we see what happens when $C'_{s'}>1$. From Eq.(\ref{AHC1}) we can find out how the Weyl nodes contribute (we call it $\Re {\sigma^{II}_{xy,s'}}$). Following the same procedure as already described 
for the case $C'_{s'}<1$, we get
\bea
&& \Re {\sigma^{II}_{xy,s'}}=\frac{e^2}{8\pi^2} \biggl[ \int^{\Lambda+Q}_{0} \frac{k_{z}dk_{z}}{C'_{s'}k_{z}+Q'_{0}}-\int^{\frac{Q'_{0}}{C'_{s'}-1}}_{\frac{Q'_{0}}{C'_{s'}+1}} dk_{z}-\nonumber \\
&& \int^{\Lambda-Q}_{\frac{Q'_{0}}{C'_{s'}-1}} \frac{k_{z}dk_{z}}{C'_{s'}k_{z}-Q'_{0}}-\int^{\frac{Q'_{0}}{C'_{s'}+1}}_{0}\frac{k_{z}dk_{z}}{Q'_{0}-C'_{s'}k_{z}}\biggr],
\eea
which can be written as,
\bea
&& \frac{\Re {\sigma^{II}_{xy,s'}}}{e^2 Q/2\pi^2} =\frac{1}{2C'_{s'}}-\frac{1}{4C'^{2}_{s'}}\biggl\{\ln(C'^{2}_{s'}\Lambda^{2}-(C'_{s'}Q+Q'_{0})^2)\nonumber\\
&& -\ln(\frac{Q'^2_{0}}{C'^2_{s'}-1})\biggr\}\frac{Q'_{0}}{Q}\nonumber\\
&& \cong \frac{1}{2C'_{s'}}+\frac{1}{4C'^{2}_{s'}}\ln(\frac{Q'^2_{0}}{C'^{2}_{s'}(C'^2_{s'}-1)\Lambda^{2}}) \frac{Q'_{0}}{Q},
\label{HallII-anti-clock}
\eea
where we have assumed that the cut off $\Lambda$ dominates over $Q$ and $Q_{0}$.

We can also study the effect of tilting the Weyl nodes in the opposite direction (clockwise). For this we have to replace $C'_{s'}$ with $-|C'_{s'}|$ in Eq.(\ref{AHC1}). Following similar algebra to what we 
have just done we find
\bea
&& \Re{\sigma^{I}_{xy,s'}}=\frac{e^2}{8\pi^2} \biggl[\int^{\frac{Q'_{0}}{1+|C'_{s'}|}}_{0} \frac{k_{z}dk_{z}}{Q'_{0}-|C'_{s'}|k_{z}}+\int^{\Lambda+Q}_{\frac{Q'_{0}}{1+|C'_{s'}|}} dk_{z}   \nonumber\\
&& -\int^{\frac{Q'_{0}}{1-|C'_{s'}|}}_{0} \frac{k_{z}dk_{z}}{|C'_{s'}|k_{z}+Q'_{0}}-\int^{\Lambda-Q}_{\frac{Q'_{0}}{1-|C'_{s'}|}} dk_{z}\biggr] 
\eea
and get,
\bea
&& \hspace{-0.7cm}\frac{\Re{\sigma^{I}_{xy,s'}}}{e^2 Q/2\pi^2}=\frac{1}{2}-\frac{1}{2}\biggl\{\hspace{-0.1cm}\frac{1}{|C'_{s'}|}+\frac{1}{2C'^{2}_{s'}}\ln\lp \frac{1-|C'_{s'}|}{1+|C'_{s'}|}\rp\hspace{-0.1cm}\biggr\}\frac{Q'_{0}}{Q},
\label{HallI-clock}
\eea 
which shows that changing the tilt from counter clockwise to clockwise changes the sign in the second term of Eq.(\ref{HallI-clock}). For a type II Weyl node we get,
\bea
&& \hspace{-0.7cm}\frac{\Re{\sigma^{II}_{xy,s'}}}{e^2 Q/2\pi^2} =\frac{1}{2|C'_{s'}|}-\frac{1}{4C'^{2}_{s'}}\ln(\frac{Q'^2_{0}}{C'^2_{s'}(C'^2_{s'}-1)\Lambda^2})\frac{Q'_{0}}{Q}
\label{HallII-clock}
\eea
where the second term carries the opposite sign from that in Eq.(\ref{HallII-anti-clock}).

In the next few subsections we will deal with different combinations of tilting of the Weyl nodes. In all these subsections $C'_{s'}$ stands for the absolute value of the tilts.

\begin{figure}[H]
\includegraphics[width=2.5in,height=3.2in, angle=270]{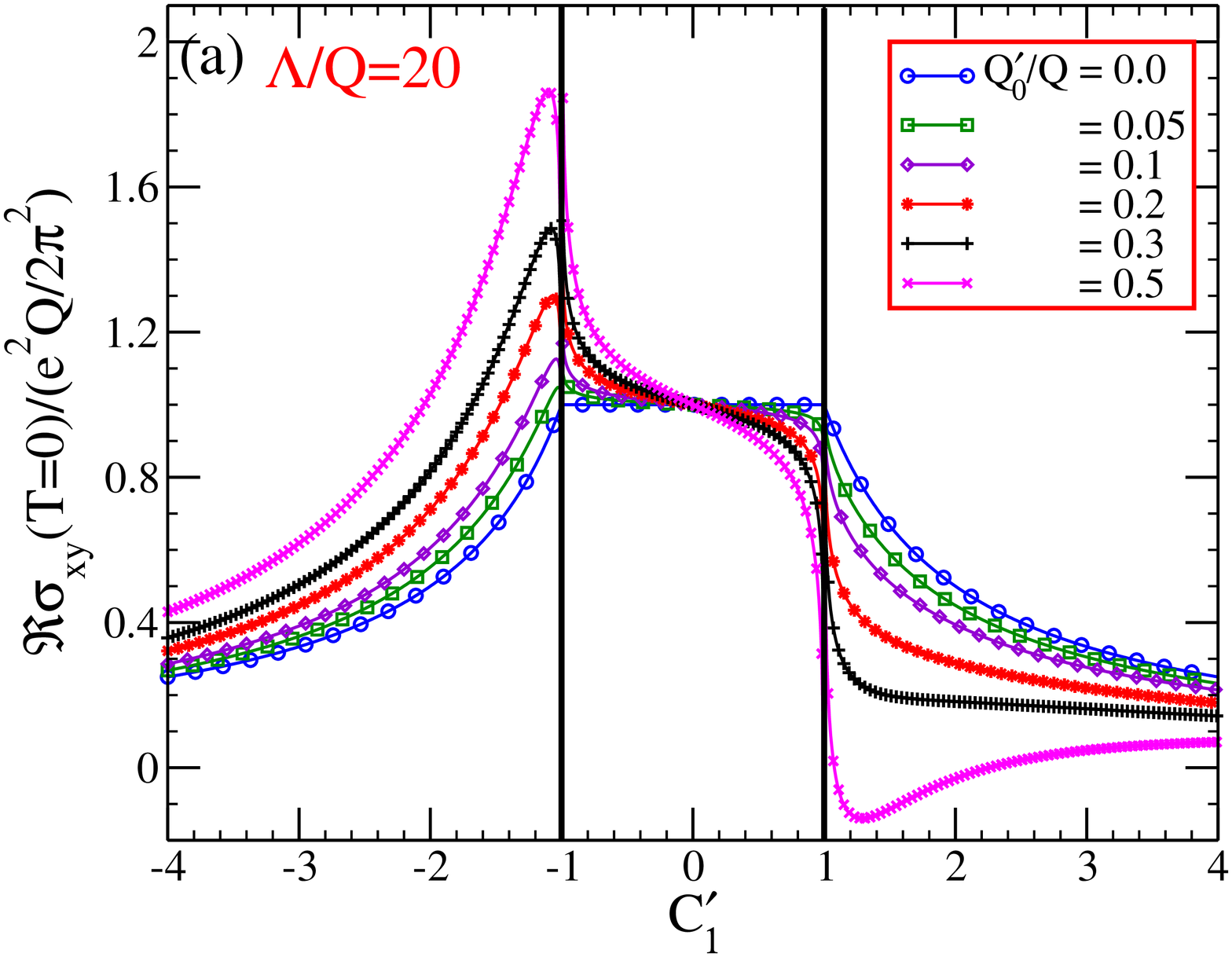} 
\includegraphics[width=2.5in,height=3.2in, angle=270]{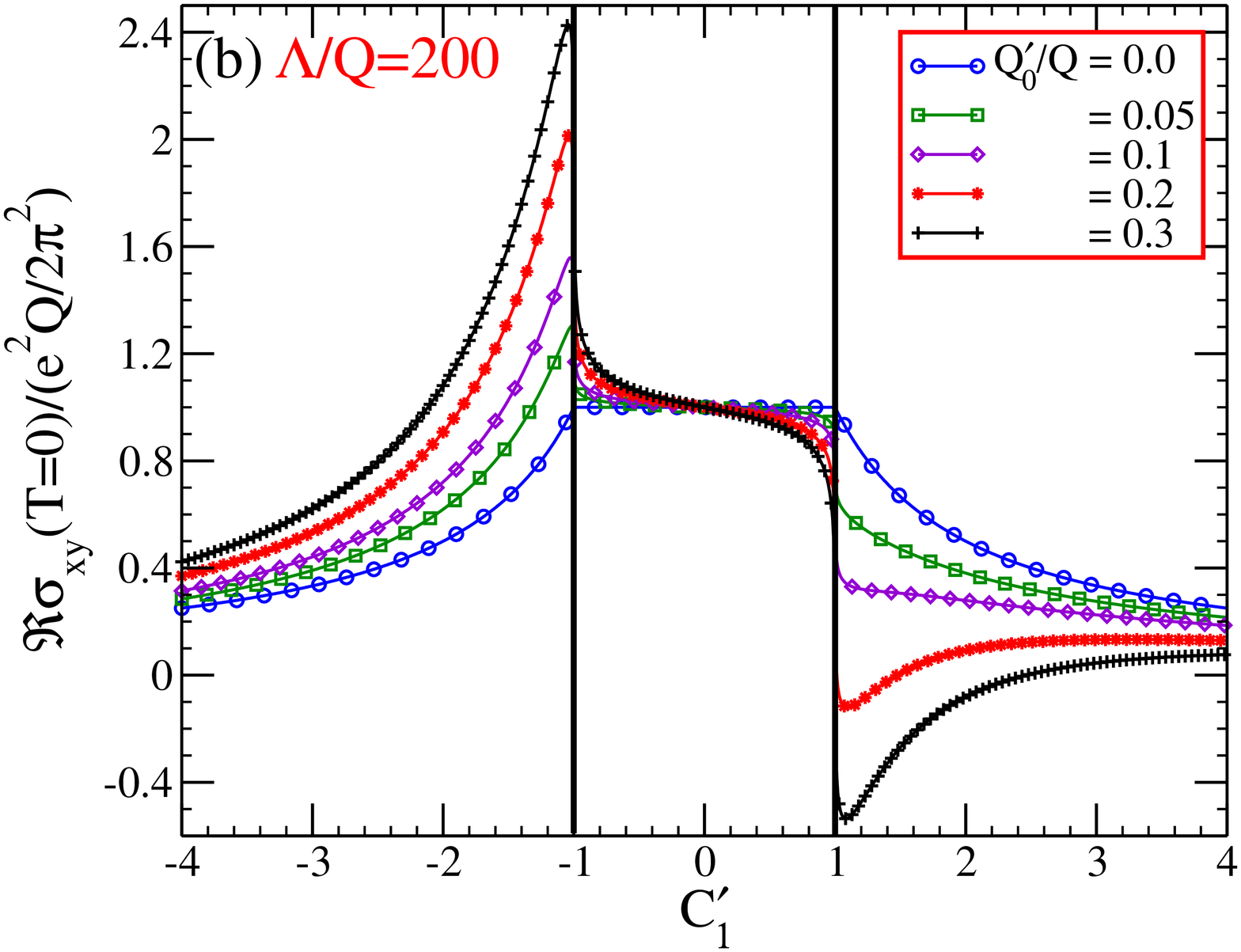} 
\caption{(Color online) We plot with dotted lines the anomalous Hall conductivity $\Re {\sigma_{xy}}$ in the units of $e^2 Q/2\pi^2$ against the amount of tilt $C'_{1}$ for different values of $Q'_{0}/Q$ 
at charge neutrality $\mu=0$. The absolute value of the amount of tilt $C'_{1}$ is same for both of the Weyl cones. This diagram covers both WSM type I as well as type II when both are tilted either 
clockwise or anti-clockwise. Blue circular dotted line is for $Q'_{0}/Q=0$, green square dotted line for 0.05, violet diamond dotted line for 0.1, red star dotted line for 0.2, black plus dotted line for 
0.3 and magenta cross dotted line for $Q'_{0}/Q=0.5$. We also show the effect of the cutoff on the anomalous conductivity $\sigma_{xy}$ by choosing two values of the cutoff, in (a) $\Lambda/Q=20$ and in 
(b) $\Lambda/Q=200$.} 
\label{Fig5}
\end{figure}

\subsection{Both the cones are tilted anti-clockwise}
\subsubsection{WSM type I}
Here we have to add $\Re {\sigma^{I}_{xy,s'}}$ for $s'=\pm$. This gives us from Eq.(\ref{HallI-anti-clock}),

\bea
&& \frac{\Re {\sigma_{xy}}}{e^2 Q/2\pi^2}=\sum_{s'=\pm}\frac{\Re {\sigma^{I}_{xy,s'}}}{e^2 Q/2\pi^2}\nonumber \\
&& =1+\frac{1}{2}\sum_{s'=\pm}\lb\frac{1}{C'_{s'}}+\frac{1}{2C'^{2}_{s'}}\ln\lp \frac{1-C'_{s'}}{1+C'_{s'}}\rp\rb\frac{Q'_{0}}{Q}.
\label{WSMI-anti-clock}
\eea
When the magnitude of the tilt are the same $C'_{1}=C'_{2}$ this reduces to,
\be
\frac{\Re {\sigma_{xy}}}{e^2 Q/2\pi^2}=1+\lb\frac{1}{C'_{1}}+\frac{1}{2C'^{2}_{1}}\ln\lp \frac{1-C'_{1}}{1+C'_{1}}\rp\rb\frac{Q'_{0}}{Q}.
\label{WSMI-anti-clock-Equal}
\ee
For both tilts clockwise we would get a minus sign between first and second term of Eq.(\ref{WSMI-anti-clock-Equal}).

\subsubsection{WSM type II}
To get the anomalous Hall conductivity we have to add $\Re {\sigma^{II}_{xy,s'}}$ for $s'=\pm$ according to Eq.(\ref{HallII-anti-clock}). This gives us,
\bea
&& \hspace{-0.4cm}\frac{\Re {\sigma_{xy}}}{e^2 Q/2\pi^2}=\sum_{s'=\pm}\frac{\Re {\sigma^{II}_{xy,s'}}}{e^2 Q/2\pi^2}\nonumber \\
&&= \sum_{s'=\pm} \biggl[\frac{1}{2C'_{s'}}+\frac{1}{4C'^{2}_{s'}}\ln(\frac{Q'^2_{0}}{C'^2_{s'}(C'^2_{s'}-1)\Lambda^2})\frac{Q'_{0}}{Q}\biggr].
\eea 
For $C'_{1}=C'_{2}$ we get
\bea
&& \frac{\Re {\sigma_{xy}}}{e^2 Q/2\pi^2}= \frac{1}{C'_{1}} + \frac{1}{2C'^2_{1}}\ln(\frac{Q'^2_{0}}{C'^2_{1}(C'^2_{1}-1)\Lambda^2})\frac{Q'_{0}}{Q}
\label{WSMII-anti-clock-Equal}
\eea
and again for both tilts clockwise there would be a minus sign between first and second term of Eq.(\ref{WSMII-anti-clock-Equal}).

In Fig.[\ref{Fig5}] we plot the anomalous Hall conductivity $\Re {\sigma_{xy}}$ in the units of $e^2 Q/2\pi^2$ against the magnitude of tilt $C'_{1}$ for different values of $Q'_{0}/Q$. Here we 
have assumed that both the cones are tilted by the same amount. We show results for both WSM type I ($C'_{1}<1$) as well as WSM type II ($C'_{1}>1$). This figure also shows the behavior of anomalous Hall 
conductivity when both nodes are tilted anti-clockwise (for positive $C'_{1}$) and clockwise (for negative $C'_{1}$). This covers all four cases discussed in this subsection. In the top frame we set the 
cut off $\Lambda/Q=20$ while in the bottom frame it is increased to 200. There is no qualitative changes introduced from the change in cut off and the quantitative differences are largest as the region 
$C'_{1}=\pm1$ is approached. This is precisely the region where the change from type I to type II occurs and there is a Lifshitz transition from a point like Fermi surface to the existence of electron and 
hole pockets at the Weyl point and our linear model is no longer realistic. These issues are further elaborated upon in Ref.[\onlinecite{Tiwari}] and [\onlinecite{Ferreiros}]. As a result here we have 
blacked out this region in Fig.(\ref{Fig5}) with thick solid vertical black lines. 

\subsection{$s'=+$ node is tilted clockwise and $s'=-$ node is tilted anti-clockwise}
\subsubsection{WSM type I}
To get the anomalous Hall conductivity in this case we have to add $\Re {\sigma^{I}_{xy,-}}$ and $\Re{\sigma^{I}_{xy,+}}$ as they appear in Eq.(\ref{HallI-anti-clock})(for $s'=-$) and (\ref{HallI-clock})
(for $s'=+$) respectively. Here $C'_{s'}$ stands for the absolute value of the tilts. This gives us,
\bea
&& \frac{\Re {\sigma_{xy}}}{e^2 Q/2\pi^2}=\frac{\Re {\sigma^{I}_{xy,+}}}{e^2 Q/2\pi^2}+\frac{\Re {\sigma^{I}_{xy,-}}}{e^2 Q/2\pi^2}\nonumber \\
&& =1-\frac{1}{2}\biggl\{\lp\frac{1}{C'_{1}}-\frac{1}{C'_{2}}\rp+\frac{1}{2C'^{2}_{1}}\ln\lp \frac{1-C'_{1}}{1+C'_{1}}\rp-\nonumber\\
&& \frac{1}{2C'^{2}_{2}}\ln\lp \frac{1-C'_{2}}{1+C'_{2}}\rp\biggr\}\frac{Q'_{0}}{Q}
\eea
Which shows that the Hall conductivity is truly universal when the absolute value of the tilts are the same, giving us,
\be
\frac{\Re {\sigma_{xy}}}{e^2 Q/2\pi^2}=1
\ee
whatever the value of $Q_{0}$ may be.

\subsubsection{WSM type II}
To get the anomalous Hall conductivity in this case we have to add $\Re{\sigma^{II}_{xy,-}}$ and $\Re{\sigma^{II}_{xy,+}}$ as they appear in Eq.(\ref{HallII-anti-clock})(for $s'=-$) and 
(\ref{HallII-clock}) (for $s'=+$) respectively. Here $C'_{s'}$ stands for the absolute value of the tilts. This gives us,
\bea
&& \frac{\Re {\sigma_{xy}}}{e^2 Q/2\pi^2}=\frac{\Re {\sigma^{II}_{xy,+}}}{e^2 Q/2\pi^2}+\frac{\Re {\sigma^{II}_{xy,-}}}{e^2 Q/2\pi^2}\nonumber \\
&& = \frac{1}{2}\lp\frac{1}{C'_{1}}+ \frac{1}{C'_{2}}\rp - \frac{1}{4}\biggl\{\frac{1}{C'^{2}_{1}}\ln(\frac{Q'^2_{0}}{C'^2_{1}(C'^2_{1}-1)\Lambda^2})\nonumber \\
&& -\frac{1}{C'^{2}_{2}}\ln(\frac{Q'^2_{0}}{C'^2_{2}(C'^2_{2}-1)\Lambda^2})\biggr\}\frac{Q'_{0}}{Q}
\eea
For same magnitude of tilt we can further simplify it to get,
\bea
&& \frac{\Re {\sigma_{xy}}}{e^2 Q/2\pi^2}=\frac{1}{C'_{1}},
\eea
again independent of $Q_{0}$.

\section{Anomalous Hall conductivity for finite chemical potential}
\label{sec:IV}

The generalization to the finite chemical potential case is straight forward and requires the mapping of $Q_{0}$ of the previous section to $s'\mu_{s'}$. The anomalous Hall conductivity for a pair of 
Weyl nodes for a general value of the distance in momentum space between the nodes ($Q$), shift in energy of the nodes ($Q_{0}$) due to the violation of inversion symmetry, for a general tilt ($t_{s'}$) 
and magnitude of this tilt $C_{s'}$ taken to be always positive is,

\bea
&& \hspace{-0.8cm}\frac{\Re {\sigma_{xy}}}{e^2/2\pi^2}\hspace{-0.1cm}=\hspace{-0.3cm}\sum_{s'=\pm}\hspace{-0.1cm}\lb\frac{Q}{2}+\frac{t_{s'}s'}{2}\biggl\{\hspace{-0.1cm}\frac{1}{C'_{s'}}+\frac{1}{2C'^{2}_{s'}}\ln\hspace{-0.1cm}\lp \frac{1-C'_{s'}}{1+C'_{s'}}\rp\hspace{-0.1cm}\biggr\}\mu'_{s'}\rb
\label{HallI-anti-clock-finiteCP}
\eea 
for $C_{s'}<1$ (type I) and
\bea
&& \hspace{-0.6cm}\frac{\Re {\sigma_{xy}}}{e^2/2\pi^2}\hspace{-0.1cm}=\hspace{-0.3cm}\sum_{s'=\pm}\hspace{-0.1cm}\lb\frac{Q}{2C'_{s'}}+\frac{t_{s'}s'}{4C'^{2}_{s'}}\ln(\frac{\mu'^2_{s'}}{C'^{2}_{s'}(C'^2_{s'}-1)\Lambda^{2}})\mu'_{s'}\rb
\label{HallII-anti-clock-finiteCP}
\eea
for $C_{s'}>1$ (type II). Here we have assumed that the tilt index $t_{s'}=\pm1$ for counterclockwise and clockwise tilt respectively irrespective of $s'$ where $s'$ is the index on the chirality of the 
Weyl nodes. 
\begin{figure}
\includegraphics[width=2.5in,height=3.2in, angle=270]{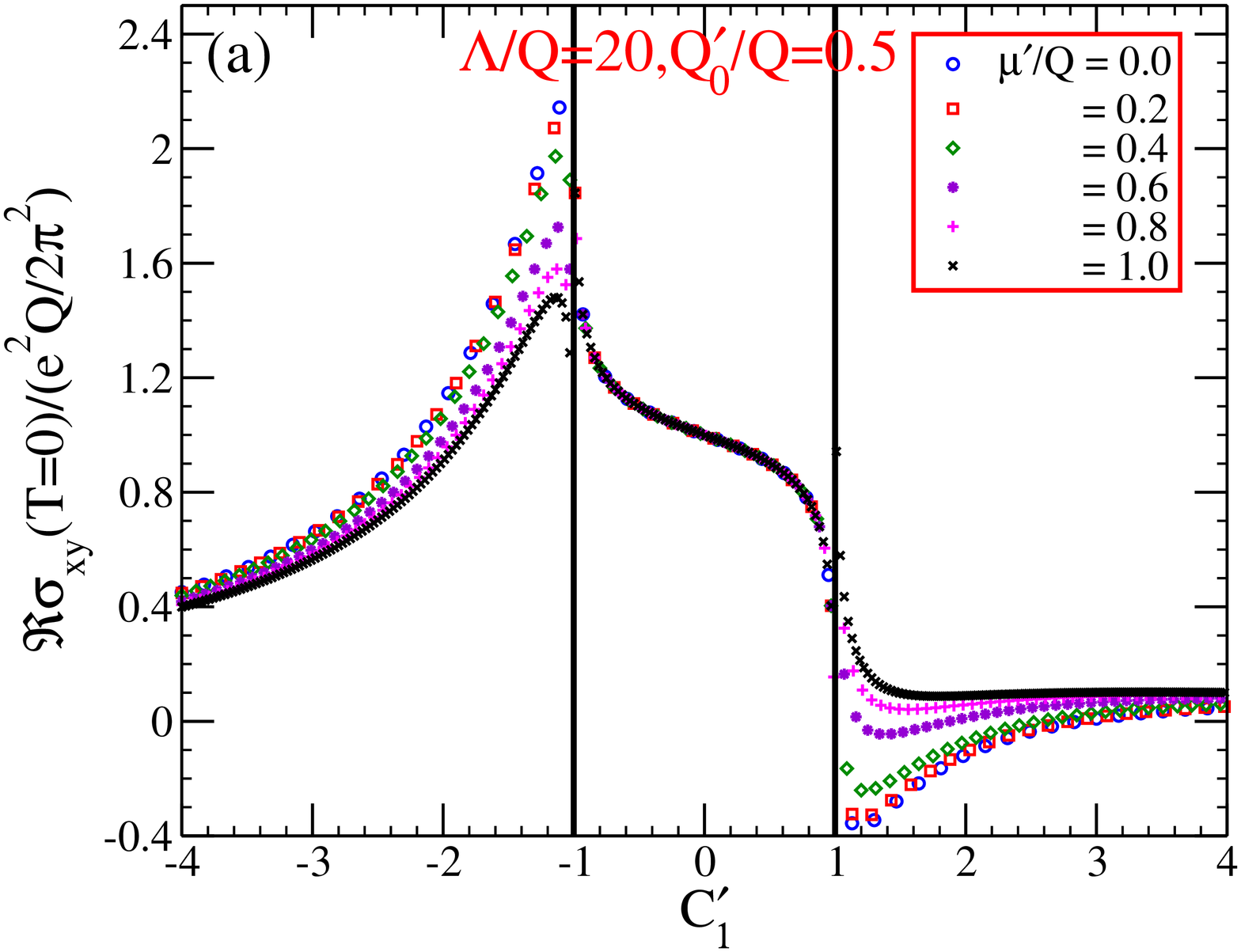} 
\includegraphics[width=2.5in,height=3.2in, angle=270]{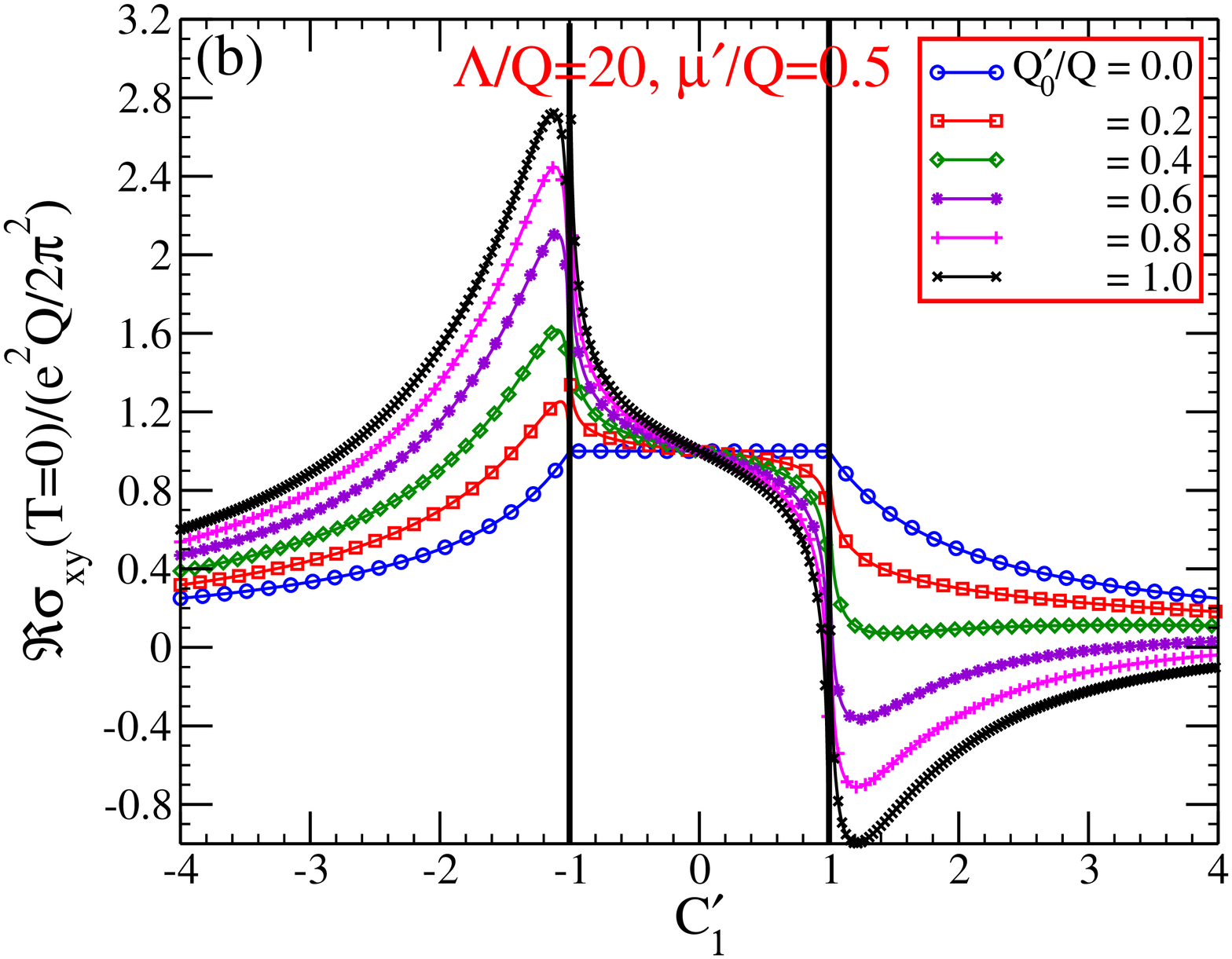} 
\caption{(Color online) We plot with dotted lines the anomalous Hall conductivity $\Re {\sigma_{xy}}$ in the units of $e^2 Q/2\pi^2$ against the amount of tilt $C'_{1}$ for the finite chemical potential
($\mu\ne0$). It is assumed that the Weyl nodes are tilted parallel to each other and the absolute value of the amount of tilt is same for both of the Weyl nodes and assumed to be $C'_{1}$. This diagram 
covers both WSM type I as well as type II when both are tilted either clockwise or anti-clockwise. In the upper panel (a) we fix $Q'_{0}/Q$ at 0.5 and generate plots with dotted lines for different 
values of the chemical potential normalized by $Q$. Blue circular dots are for $\mu'/Q=0$, red square dots for 0.2, green diamond dots for 0.4, violet star dots for 0.6, magenta plus dots for 0.8 and 
black cross dots for $\mu'/Q=1$. In the bottom panel (b) we fix $\mu'/Q$ at 0.5 and plot for different values of $Q'_{0}/Q$ namely blue circular dotted line for $Q'_{0}/Q=0$, red square dotted line for 
0.2, green diamond dotted line for 0.4, violet star dotted line for 0.6, magenta plus dotted line for 0.8 and black cross dotted line for $Q'_{0}/Q=1$.}
\label{Fig6}
\end{figure}

These equations are the central results of this work. The first term in these equations is directly proportional to the distance in momentum space of the two nodes $Q$ and does not depend on doping $\mu\ne0$ and on the energy shift due to inversion 
symmetry breaking $Q_{0}$. The second term is a correction due to $\mu$ and $Q_{0}$ but does not depend on $Q$ in the approximation used here namely that the cut off $\Lambda$ is much larger than both 
$Q$ and $Q_{0}$ in momentum unit. While for simplicity we have assumed that the tilt of the negative chirality node is counterclockwise, changing to clockwise simply changes the sign of this contribution 
in the second term of both Eq.(\ref{HallI-anti-clock-finiteCP}) and (\ref{HallII-anti-clock-finiteCP}) and will not be explicitly treated beyond this comment. It is instructive to write down separately 
the case of the positive chirality node tilted to the left (counterclockwise) and to the right (clockwise). In the first case we have
\bea
&& \hspace{-0.8cm}\frac{\Re {\sigma_{xy}}}{e^2/2\pi^2}\hspace{-0.1cm}=Q+\frac{1}{2}\hspace{-0.1cm}\sum_{s'=\pm}\hspace{-0.1cm}s'\biggl\{\hspace{-0.1cm}\frac{1}{C'_{s'}}+\frac{1}{2C'^{2}_{s'}}\ln\hspace{-0.1cm}\lp \frac{1-C'_{s'}}{1+C'_{s'}}\rp\hspace{-0.1cm}\biggr\}\mu'_{s'}
\label{HallI-anti-clock-finiteCP-positive}
\eea 
for $C_{s'}<1$ (type I) and
\bea
&& \hspace{-0.6cm}\frac{\Re {\sigma_{xy}}}{e^2/2\pi^2}\hspace{-0.1cm}=\hspace{-0.1cm}Q\hspace{-0.2cm}\sum_{s'=\pm}\hspace{-0.1cm}\frac{1}{2C'_{s'}}+\hspace{-0.3cm}\sum_{s'=\pm}\hspace{-0.1cm}\frac{s'}{4C'^{2}_{s'}}\ln(\frac{\mu'^2_{s'}}{C'^{2}_{s'}(C'^2_{s'}-1)\Lambda^{2}})\mu'_{s'}
\label{HallII-anti-clock-finiteCP-positive}
\eea
for $C_{s'}>1$ (type II). We have not assumed that the magnitude of the tilts is the same for each Weyl node but we have written the formula for the case that, either both are type I or both type II. 
For the second case when the negative chirality cone is tilted counterclockwise but the positive chirality cone is clockwise we have instead
\bea
&& \hspace{-0.8cm}\frac{\Re {\sigma_{xy}}}{e^2/2\pi^2}\hspace{-0.1cm}=Q-\frac{1}{2}\hspace{-0.1cm}\sum_{s'=\pm}\hspace{-0.1cm}\biggl\{\hspace{-0.1cm}\frac{1}{C'_{s'}}+\frac{1}{2C'^{2}_{s'}}\ln\hspace{-0.1cm}\lp \frac{1-C'_{s'}}{1+C'_{s'}}\rp\hspace{-0.1cm}\biggr\}\mu'_{s'}
\label{HallI-anti-clock-finiteCP-PN}
\eea 
for $C_{s'}<1$ (type I) and
\bea
&& \hspace{-0.6cm}\frac{\Re {\sigma_{xy}}}{e^2/2\pi^2}\hspace{-0.1cm}=\hspace{-0.1cm}Q\hspace{-0.2cm}\sum_{s'=\pm}\hspace{-0.1cm}\frac{1}{2C'_{s'}}-\hspace{-0.3cm}\sum_{s'=\pm}\hspace{-0.1cm}\frac{1}{4C'^{2}_{s'}}\ln(\frac{\mu'^2_{s'}}{C'^{2}_{s'}(C'^2_{s'}-1)\Lambda^{2}})\mu'_{s'}
\label{HallII-anti-clock-finiteCP-PN}
\eea
for $C_{s'}>1$ (type II). These equations simplify if we assume the magnitude of the tilt to be the same for each of the two nodes. Denoting the tilt simply by $C$ we get for $C<1$,
\be
\frac{\Re{\sigma_{xy}}}{e^2/2\pi^2}=Q+\lb\frac{1}{C}+\frac{1}{2C^{2}}\ln\lp\frac{1-C}{1+C}\rp\rb Q'_{0},
\label{WSMI-parallel-equalC}
\ee 
which is independent of the chemical potential and for $C>1$
\bea
&& \frac{\Re{\sigma_{xy}}}{e^2/2\pi^2}=\frac{Q}{C}+\frac{1}{2C^{2}}\biggl[\ln\lp \frac{|\mu'^2-Q'^2_{0}|}{C^{2}(C^2-1)\Lambda^{2}}\rp Q'_{0}+ \nonumber\\ 
&& \ln\biggl|\frac{\mu'+Q'_{0}}{\mu'-Q'_{0}}\biggr|\mu'\biggr], 
\label{WSMII-parallel-equalC}
\eea
which applies to parallel tilts. For opposite tilting of the two nodes we get instead for $C<1$
\be
\frac{\Re{\sigma_{xy}}}{e^2/2\pi^2}=Q-\lb\frac{1}{C}+\frac{1}{2C^{2}}\ln\lp\frac{1-C}{1+C}\rp\rb \mu',
\label{WSMI-antiparallel-equalC}
\ee 
which is now independent of the inversion symmetry breaking energy shift $Q_{0}$. For $C>1$
\bea
&& \frac{\Re{\sigma_{xy}}}{e^2/2\pi^2}=\frac{Q}{C}-\frac{1}{2C^{2}}\biggl[\ln\lp \frac{|\mu'^2-Q'^2_{0}|}{C^{2}(C^2-1)\Lambda^{2}} \rp \mu' + \nonumber \\ 
&& \ln\biggl|\frac{\mu'+Q'_{0}}{\mu'-Q'_{0}}\biggr|Q'_{0}\biggr].
\label{WSMII-antiparallel-equalC}
\eea
Comparing these two cases more closely we note that the sign of the correction term to the usual topological term proportional to the distance in momentum space between the two Weyl nodes is opposite 
in Eq.(\ref{WSMI-parallel-equalC}) and (\ref{WSMII-parallel-equalC}) to that in (\ref{WSMI-antiparallel-equalC}) and (\ref{WSMII-antiparallel-equalC}). Also $Q_{0}$ and $\mu$ have switched role. For 
$C<1$ (type I) the corrections for finite $Q_{0}$ in Eq.(\ref{WSMI-parallel-equalC}) and for finite $\mu$ in Eq.(\ref{WSMI-antiparallel-equalC}) is linear. This is no longer the case for $C>1$ (type II) 
for which both $Q_{0}$ and $\mu$ terms arise and the contribution of such terms to the anomalous Hall is no longer linear in $Q_{0}$ or in $\mu$ but involves logarithmic correction. In Fig.[\ref{Fig6}] 
we plot the DC Hall conductivity $\Re {\sigma_{xy}}$ in units of $e^2 Q/2\pi^2$ as a function of the tilt. The figure is drawn assuming that the tilts of the two Weyl nodes are always parallel so that 
they violate tilt inversion symmetry. For positive $C$ both tilts are counterclockwise while for negative value of $C$ both tilts are clockwise. Eq.(\ref{WSMI-parallel-equalC}) applies for $|C'_{1}|=C<1$. 
$\Re {\sigma_{xy}}$ in the chosen units is $ 1+\lb\frac{1}{C}+\frac{1}{2C^{2}}\ln\lp\frac{1-C}{1+C}\rp\rb 0.5$ independent of the chemical potential so there is a single curve in Fig.[\ref{Fig6}(a)]. 
For positive $C'_{1}=C$ the correction to one is negative and in fact goes like $-C/6$ for small $C$ while for $C'_{1}$ negative the correction is additive. In the overtilted case (type II) this simplicity 
no longer applies and it is Eq.(\ref{WSMII-parallel-equalC}) and (\ref{WSMII-antiparallel-equalC}) that are relevant. There is now a dependence on both $Q_{0}$ and $\mu$ which is no longer linear and the 
cutoff $\Lambda$ enters. The dependence on cutoff is, as we saw in Fig.[\ref{Fig5}], most pronounced as $|C'_{1}|\to 1$ and the Lifshitz transition is approached. In this case our calculations are no 
longer expected to be quantitatively accurate and this is manifested in a logarithmic singularity. This logarithmic singularity is also seen as $|C'_{1}|\to 1$ from below. In the lower frame we have fixed 
the value of $\mu$ at $\mu'/Q=0.5$ and vary $Q'_{0}/Q$ but of course we know from Eq.(\ref{WSMI-parallel-equalC}) that any other choice of $\mu$ would give the same result. Starting with the region 
$|C'_{1}|<1$ the family of curves is given by $ 1+\lb\frac{1}{C}+\frac{1}{2C^{2}}\ln\lp\frac{1-C}{1+C}\rp\rb \frac{Q'_{0}}{Q}$ and so we get variations with changing value of $Q'_{0}/Q$. Note in particular 
that for $Q'_{0}/Q=0$ we get a constant equal to 1 in our units, independent of the magnitude of the tilt. For finite $Q'_{0}/Q$ but small value of tilt we get $1-\frac{C}{3}\lp\frac{Q'_{0}}{Q}\rp$ with 
the slope out of $C'_{1}=0$ linearly increasing with $Q'_{0}/Q$. If we had instead considered the tilts in the Weyl pair of nodes to be oppositely directed (respect tilt inversion symmetry) rather than 
be parallel Eq.(\ref{WSMI-antiparallel-equalC}) would apply instead of Eq.(\ref{WSMI-parallel-equalC}) so the same figure would describe the Hall conductivity with the role of $Q_{0}$ and $\mu$ interchanged 
and a sign change on the tilt. Such a set of results are given in Fig.[3] of Ref.[\onlinecite{Tiwari}] where the $Q_{0}=0$ case was considered.

Finally, we comment briefly on the effect the chiral anomaly can have on the anomalous Hall conductivity. A non-zero value of the dot product $\textbf{E.B}$ of electric $\textbf{E}$ and magnetic 
$\textbf{B}$ field, will pump charge from one Weyl node to the other\cite{Ashby,Hosur1,Xiao1} changing their chemical potential at charge neutrality and zero tilt by $s'\mu_{p}$ with
\be
\mu_{p}=\lb \frac{3e^2\hbar v^3}{2}(\textbf{E.B})\tau_{inter}\rb^{1/3}
\label{chiral}
\ee
where $s'$ is the chirality and $\tau_{inter}$ is an intervalley relaxation time. This contribution is equivalent to a finite $Q_{0}$ value of magnitude (given in Eq.(\ref{chiral})) $\mu_{p}$. For small 
tilt $C<<1$ and the parallel tilt case we need not correct Eq.(\ref{chiral}) for tilt to lowest order and the anomalous Hall conductivity, in our units, will be equal to 
$\lp 1- \frac{C}{6}\frac{\mu_{p}}{v_{z}Q}\rp$. For $\mu_{p}\sim 10$meV\cite{Ashby}, $v_{z}\sim 3\times10^5$m/s\cite{Huang} and $Q\sim 0.08\AA$\cite{Xu}, $\mu_{p}/\hbar v_{z}Q$ is of the order 0.06 which 
is small.

\section{Nernst and thermal Hall}
\label{sec:V}

In a recent paper Ferreiros et. al. [\onlinecite{Ferreiros}] have discussed the effect of tilt on the Nernst and thermal Hall coefficient in Weyl semimetals. Here we generalize their work to 
noncentrosymmetric materials. Both Nernst ($\alpha_{xy}$) and thermal Hall conductivity ($\kappa_{xy}$) depend on the Hall conductivity $\Re{\sigma_{xy}}$ of the previous sections which from here on 
we will denote simply by $\sigma_{xy}$ suppressing the real part index. They are \cite{Ferreiros}
\be
\alpha_{xy}= \frac{\pi^2 k^2_{B}T}{3e} \frac{d\sigma_{xy}}{d\mu}
\label{Nernst-coeff}
\ee
and 
\be
\kappa_{xy}= \frac{\pi^2 k^2_{B}T}{3e^2}\sigma_{xy}
\label{thermal-cond-coeff}
\ee
with $k_{B}$ the Boltzmann constant, $e$ the electronic charge and $T$ the temperature.  The thermal Hall coefficient in Eq.(\ref{thermal-cond-coeff}) is simply proportional to the Hall conductivity and 
is therefore given by Eq.(\ref{WSMI-parallel-equalC}) to (\ref{WSMII-antiparallel-equalC}) multiplied by $\frac{\pi^2 k^2_{B} T}{3e^2}$. As we have discussed in the previous section for type I Weyl, the 
Hall conductivity has a logarithmic singularity (Eq.(\ref{WSMI-parallel-equalC}) and (\ref{WSMII-parallel-equalC})) as the Lifshitz transition at $C=1$ is approached. Our results are no longer 
expected to be quantitatively accurate in this region. For type II there is a related logarithmic singularity at $\mu=Q_{0}$. This is expected since the effective chemical potential of the negative 
chirality node is zero at $\mu=Q_{0}$ and we are again probing the nodal region for which our continuum Hamiltonian requires modification. Here we have introduced a cut off in momentum $\Lambda$ to regularize 
our results. In view of this limitation we will consider only the case $Q_{0}/\mu<1$ and $\mu/Q_{0}<1$. Both logarithms can be expanded in these two limiting cases. Eq.(\ref{WSMII-parallel-equalC}) for 
parallel tilt then takes the form,
\bea
\label{Nernst1}
&& \frac{\Re{\sigma_{xy}}}{e^2/2\pi^2}\cong\frac{Q}{C}+\frac{1}{2C^{2}}\biggl[\lf\ln\lp \frac{\mu'^2}{C^{2}(C^2-1)\Lambda^{2}}\rp +2\rf Q'_{0} \nonumber\\
&& -\frac{1}{3} \lp\frac{Q'_{0}}{\mu'}\rp^3 \mu'\biggr], \hspace{3.0cm}\text{for} \frac{Q'_{0}}{\mu'}<1
\eea
\bea
&& \cong\frac{Q}{C}+\frac{1}{2C^{2}} \biggl\{\ln\lp \frac{Q'^2_{0}}{C^{2}(C^{2}-1)\Lambda^{2}}\rp + \lp\frac{\mu'}{Q'_{0}}\rp^2 + \nonumber \\ 
&& \frac{1}{6} \lp\frac{\mu'}{Q'_{0}}\rp^4\biggr\} Q'_{0}, \hspace{3.2cm}\text{for} \frac{\mu'}{Q'_{0}}<1.
\label{Nernst2}
\eea
This means that for $C<1$ (type I) (see Eq.(\ref{WSMI-parallel-equalC})) the Nernst coefficient is zero for parallel tilts and for $C>1$ it is,
\bea
&& \hspace{-0.4cm}\alpha_{xy}= \frac{e k^2_{B}T}{6} \lb \frac{Q'_{0}}{\mu'}+ \frac{1}{3} \lp \frac{Q'_{0}}{\mu'} \rp^3 \rb \frac{1}{C^{2}},~~\text{for} \frac{Q'_{0}}{\mu'}<1 \\
&& ~~=\frac{e k^2_{B}T}{6} \lb \frac{\mu'}{Q'_{0}} + \frac{2}{3} \lp\frac{\mu'}{Q'_{0}}\rp^3\rb \frac{1}{C^{2}},~\text{for} \frac{\mu'}{Q'_{0}}<1.
\eea
For $Q_{0}=0$ we recover the result found in Ref.[\onlinecite{Ferreiros}] that $\alpha_{xy}=0$. At finite $Q_{0}$ with $Q_{0}/\mu<1$ the Nernst coefficient is finite with leading order proportional to 
$Q_{0}/\mu$ and first correction of order $Q^3_{0}/\mu^3$. For $\mu/Q_{0}<1$, $\alpha_{xy}$ is also finite and to leading order in $\mu/Q_{0}$, is proportional to $\mu$ and inversely proportional to 
$Q_{0}$.

For oppositely tilted Weyl nodes (tilt inversion symmetry applies) we get for type I
\be
\alpha_{xy}=-\frac{e k^2_{B}T}{6C} \lb 1+ \frac{1}{2C} \ln \biggr| \frac{1-C}{1+C} \biggl| \rb,
\ee
which in the limit of the tilt going to zero, gives $e k^2_{B}TC/18$. This has the opposite sign to that in Ref.[\onlinecite{Ferreiros}] because we are dealing here with a counterclockwise tilt on 
the negative chirality node with the tilt of the positive chirality node clockwise. For type II a similar set of equations to Eq.(\ref{Nernst1}) and (\ref{Nernst2}) are obtained from 
Eq.(\ref{WSMII-antiparallel-equalC}) by changing the overall sign of the second term in these equations and switching the variable $Q_{0}$ and $\mu$. For $Q_{0}/\mu<1$
\bea
&& \hspace{-0.5cm}\alpha_{xy}= -\frac{e k^2_{B}T}{12 C^2} \lb \ln\lp \frac{\mu'^2}{C^2(C^2-1)\Lambda^2}\rp +2 -\lp\frac{Q'_{0}}{\mu'}\rp^2\rb,
\eea
which in the $Q_{0}=0$ limit gives
\be
\alpha_{xy}\cong -\frac{e k^2_{B}T}{6 C^2} \lb \ln \biggl|\frac{\mu'}{C\Lambda \sqrt{C^2-1}}\biggr| +1 \rb.
\ee
This agrees with Ref.[\onlinecite{Ferreiros}]. Here we have an additive correction for finite $Q_{0}$ equal to $\frac{e k^2_{B}T}{12 C^{2}}\lp\frac{Q'_{0}}{\mu'}\rp^2 $. For the opposite limit 
$\mu/Q_{0}<1$ to leading order
\be
\alpha_{xy}\cong -\frac{e k^2_{B}T}{6 C^2} \lb \ln \biggl|\frac{Q'_{0}}{C \Lambda \sqrt{C^2-1}}\biggr| +1 -\frac{1}{2} \lp\frac{\mu'}{Q'_{0}}\rp^2\rb
\ee
which is, in the leading order, independent of chemical potential $\mu'$.

\section{Discussion and conclusions}
\label{sec:VI}

We have studied the anomalous DC Hall conductivity ($\sigma_{xy}$) in a continuum Dirac model Hamiltonian which additionally includes a tilt ($C$) as well as time reversal (TR) and inversion (I) symmetry 
breaking terms. These last two terms lift the two fold degeneracy of the Dirac cone creating a pair of Weyl nodes of opposite chirality ($s'=\pm$). For broken TR symmetry the Weyl nodes are displaced 
in momentum space by $-s'\bf{Q}$ while for inversion symmetry breaking the displacement is in energy $-s'Q_{0}$. We employ Kubo formula to calculate the Hall conductivity and derive simple analytic 
expressions for $\sigma_{xy}$ which depend on tilt, $Q$ and $Q_{0}$ as well as on chemical potential $\mu$. For the case of charge neutrality and $Q_{0}=0$ (centrosymmetric case) we recover the known 
result when the magnitude of the tilt is taken to be the same on each nodes. For type I Weyl ($C<1$) $\sigma_{xy}=e^2 Q/2\pi^2$ and for type II ($C>1$) $\sigma_{xy}=e^2 Q/2\pi^2 C$. Both are proportional 
to the momentum space displacement $Q$. For type I the result is universal independent of tilt while for type II it is inversely proportional to $C$. Adding finite $\mu$ and/or finite $Q_{0}$ 
(noncentrosymmetric case) does not change in any way the contribution proportional to $Q$ but adds a correction which depends on $C$, $Q_{0}$ and $\mu$ but not on $Q$. The relative orientation of the 
tilts affects this contribution. It matters whether the tilts are parallel to each other which violates tilt inversion symmetry or they oppose each other which respect tilt inversion symmetry. For 
centrosymmetric Weyl with parallel tilts we recover the known result that there is no correction for finite $\mu$ while for oppositely oriented tilts it is proportional to $\mu$ for type I with a further 
logarithmic correction dependent on a momentum cut off $\Lambda$ multiplying $\mu$ for type II which makes this contribution non linear. For the general case the analytic expressions describing the $\mu$, 
$Q_{0}$ correction has the same general form for both tilt configuration except that the two variables $Q_{0}$ and $\mu$ are interchanged and the overall sign of this contribution is reversed. For 
oppositely tilted Weyl cones (tilt inversion symmetry is respected) and type I the correction to $e^2 Q/2\pi^2$ contribution is linear in $\mu$ and independent of $Q_{0}$ with coefficient dependent only 
on tilt magnitude. For parallel tilted Weyl this contribution is now linear in $Q_{0}$ and completely independent of $\mu$. For type II both variables $Q_{0}$ and $\mu$ feature in the correction term 
which is non linear and contains the momentum cut off $\Lambda$ on the $k_{z}$ integration as recorded in Eq. (\ref{WSMII-parallel-equalC}) and Eq.(\ref{WSMII-antiparallel-equalC}). Expansions for 
$\frac{\Re{\sigma_{xy}}}{e^2/2\pi^2}$ valid for $\mu/Q_{0}<1$ and for $Q_{0}/\mu<1$ are derived and given in Eq.(\ref{Nernst1}) and (\ref{Nernst2}) for parallel tilts. Equivalent equations for oppositely 
directed tilts can be obtained from these through an overall change in sign and switching the role of $Q_{0}$ and $\mu$.

We have also considered the effect of broken inversion symmetry on the Nernst effect and on the thermal Hall conductivity. Recently Ferreiros et. al. [\onlinecite{Ferreiros}] have considered these 
transport coefficients when only time reversal symmetry is violated. Our results properly reduce to those of Ref.[\onlinecite{Ferreiros}] when we set $Q_{0}=0$. For finite $Q_{0}$ we find corrections. 
The thermal Hall follows directly from the results that we have just summarized for the Hall conductivity after multiplying by $\pi^2 k^2_{B} T/ 3e^2$. The Nernst requires taking a derivative of 
$\sigma_{xy}$ with respect to the chemical potential. For parallel tilts we find that the Nernst effect remains zero in type I Weyl but it is no longer zero for type II with leading correction instead 
of order $Q_{0}/\mu$ for $Q_{0}/\mu<1$ and $\mu/Q_{0}$ for $\mu/Q_{0}<1$. For oppositely oriented tilts there is no correction to the Nernst coefficient due to $Q_{0}$ for type I, but for type II there 
is a correction of order $Q^{2}_{0}/\mu^2$ for $Q_{0}/\mu<1$. For $\mu/Q_{0}<1$, to leading order $\alpha_{xy}$ depends logarithmically on $Q_{0}$ and on the momentum cut off $\Lambda$ but not on $\mu$.
A first correction is of order $\lp \mu/Q_{0}\rp^2$.

\subsection*{Acknowledgments}
Work supported in part by the Natural Sciences and Engineering Research Council of Canada (NSERC)(Canada) and by the Canadian Institute for Advanced Research (CIFAR)(Canada). We thanks A.A.Burkov and 
D. Xiao for enlightening discussions.

\end{document}